\documentclass[useAMS,usenatbib]{mn2e}
\usepackage{graphicx} 
\usepackage{subcaption}
\usepackage{amsmath} 
\usepackage{xcolor}
\usepackage{float}
\usepackage{caption}
\usepackage{placeins}
\usepackage{lipsum}
\usepackage{multicol}
\usepackage{soul}
\usepackage{amssymb}

\def\be{\begin{equation}} 
\def\ee{\end{equation}}

\def\msun{{\Msun}}

\def\HI{\hbox{H~$\scriptstyle\rm I\ $}} 
\def\HII{\hbox{H~$\scriptstyle\rm II\ $}} 
 
\def\HeII{\hbox{He~$\scriptstyle\rm II\ $}} 
\def\HeIII{\hbox{He~$\scriptstyle\rm III\ $}}

\def\gsim{\lower.5ex\hbox{\gtsima}} 
\def\lsim{\lower.5ex\hbox{\ltsima}} \def\gtsima{$\; \buildrel > \over \sim \;$} \def\ltsima{$\; \buildrel < \over \sim \;$} \def\prosima{$\; 
\buildrel \propto \over \sim \;$} \def\gsim{\lower.5ex\hbox{\gtsima}} 
\def\lsim{\lower.5ex\hbox{\ltsima}} 
\def\simgt{\lower.5ex\hbox{\gtsima}} 
\def\simlt{\lower.5ex\hbox{\ltsima}} 
\def\simpr{\lower.5ex\hbox{\prosima}} \def\la{\lsim} \def\ga{\gsim} 
  
 \def\gtsima{$\; \buildrel > \over \sim \;$} 
\def\ltsima{$\; \buildrel < \over \sim \;$} 
\def\gsim{\lower.5ex\hbox{\gtsima}} 
\def\lsim{\lower.5ex\hbox{\ltsima}} 
\def\simgt{\lower.5ex\hbox{\gtsima}} 
\def\simlt{\lower.5ex\hbox{\ltsima}} 
\def\simpr{\lower.5ex\hbox{\prosima}} 
\def\la{\lsim} 
\def\mbh{\rm M_{bh}}
\def\ga{\gsim}

\def\msun{\,{\rm \Msun}}

\def\E3{{\cal E}_{\rm g}^{III}}

\def\Msun{\rm M_\odot}

\def\mstar{\rm M_*}
\def\mh{\rm M_h}
\def\Z*{Z_*}
\def\L*{L_*}
\def\muv{\rm M_{UV}}
\def\luv{\rm L_{UV}}

\def\fesc{f_{\rm esc}}
\def\fescsf{\langle f_{\rm esc}^{\rm sf} \rangle}
\def\fescbh{f_{\rm esc}^{\rm bh}}

\def\myr{\rm Myr}
\def\vvir{\rm V_{vir}}
\def \mh {\rm M_h}

\title[Reionization with galaxies and AGN]{Reionization with galaxies and active galactic nuclei} 
\author[Dayal et. al.]{Pratika Dayal$^{1}$\thanks{p.dayal@rug.nl}, Marta Volonteri$^2$, Tirthankar Roy Choudhury$^3$, 
\newauthor Raffaella Schneider$^{4,5,6}$, Maxime Trebitsch$^{2,7,8}$, Nickolay Y. Gnedin$^{9,10,11}$, 
\newauthor Hakim Atek$^{2}$, Michaela Hirschmann$^{12}$, Amy Reines$^{13}$\\ 
$^{1}$Kapteyn Astronomical Institute, University of Groningen, P.O. Box 800, 9700 AV Groningen, The Netherlands \\
$^{2}$Institut d'Astrophysique de Paris, Sorbonne Universite, CNRS, UMR 7095, 98 bis bd Arago, 75014 Paris, France\\
$^{3}$National Centre for Radio Astrophysics, Tata Institute of Fundamental Research, Pune 411007, India\\
$^{4}$Dipartimento di Fisica, ``Sapienza'' Universit$\grave{a}$ di Roma, Piazzale Aldo Moro 5, 00185 Roma, Italy \\
$^5$INAF/Osservatorio Astronomico di Roma, Via di Frascati 33, 00040 Monte Porzio Catone, Italy \\
$^6$INFN, Sezione Roma1, Dipartimento di Fisica, ``Sapienza'' Universit$\grave{a}$ di Roma, Piazzale Aldo Moro 5, 00185, Roma, Italy \\
$^7$Max-Planck-Institut f{\"u}r Astronomie, K{\"o}nigstuhl 17, 69117 Heidelberg, Germany \\
$^8$Zentrum f{\"u}r Astronomie der Universit{\"a}t Heidelberg, Institut f{\"u}r Theoretische Astrophysik, Albert-Ueberle-Str. 2, 69120 Heidelberg, Germany \\
$^9$Particle Astrophysics Center, Fermi National Accelerator Laboratory, Batavia, IL 60510, USA \\
$^{10}$Kavli Institute for Cosmological Physics, The University of Chicago, Chicago, IL 60637, USA \\
$^{11}$Department of Astronomy \& Astrophysics, The University of Chicago, Chicago, IL 60637, USA \\
$^{12}$DARK, Niels Bohr Institute, University of Copenhagen, Lyngbyvej 2, DK-2100 Copenhagen, Denmark \\
$^{13}$eXtreme Gravity Institute, Department of Physics, Montana State University, Bozeman, MT 59717, USA
}

\begin{document} 
 
 \date{} 

\maketitle

\begin{abstract}

In this work we investigate the properties of the sources that reionized the intergalactic medium (IGM) in the high-redshift Universe. Using a semi-analytical model aimed at reproducing galaxies and black holes in the first $\sim 1.5$~Gyr of the Universe, we revisit the relative role of star formation and black hole accretion in producing ionizing photons that can escape into the IGM. Both star formation and black hole accretion are regulated by supernova feedback, resulting in black hole accretion being stunted in low-mass halos. We explore a wide range of combinations for the escape fraction of ionizing photons (redshift-dependent, constant and scaling with stellar mass) from both star formation ($\fescsf$) and AGN ($\fescbh$) to find: {\it (i)} the ionizing budget is dominated by stellar radiation from low stellar mass ($M_*<10^9 \msun$ ) galaxies at $z>6$ with the AGN contribution (driven by $M_{bh}>10^6 \msun$ black holes in $M_* \gsim 10^9\msun$ galaxies) dominating at lower redshifts; {\it (ii)} AGN only contribute $10-25\%$ to the cumulative ionizing emissivity by $z=4$ for the models that match the observed reionization constraints; {\it (iii)} if the stellar mass dependence of $\fescsf$ is shallower than $\fescbh$, at $z<7$ a {\it transition stellar mass} exists above which AGN dominate the escaping ionizing photon production rate; {\it (iv)} the transition stellar mass decreases with decreasing redshift. While AGN dominate the escaping emissivity above the knee of the stellar mass function at $z \sim 6.8$, they take-over at stellar masses that are a tenth of the knee mass by $z=4$.

\end{abstract}


% *************************************************************
\section{Introduction}
\label{intro}
% *************************************************************
The epoch of (hydrogen) reionization (EoR) begins when the first stars start producing neutral hydrogen (\HI) ionizing photons and carving out ionized regions in the intergalactic medium (IGM). In the simplest picture, the EoR starts with the formation of the first metal-free (population III; PopIII) stars at $z \lsim 30$, with the key sources gradually shifting to larger metal-enriched halos, powered by population II (PopII) stars and accreting black holes.
However, this picture is complicated by the fact that the progress and sources of reionization depend on a number of (poorly constrained) parameters including the minimum halo mass of star-forming galaxies, the star formation/black hole accretion rates, the escape fraction ($f_{\rm esc}$) of \HI ionizing photons from the galactic environment, the impact of the reionization ultra-violet background (UVB) on the gas content of low-mass halos and the clumping factor of the IGM \citep[see e.g.][]{dayal2018}. 

Observationally, a number of works have used a variety of data-sets and trends - e.g. the UV luminosity density, the faint-end slope of the Lyman Break Galaxy (LBG) luminosity function, $f_{\rm esc}$ increasing with bluer UV slopes and the abundance and luminosity distribution of galaxies -  to conclude that star formation in low-mass galaxies with an absolute magnitude $\muv \simeq -10$ to  $-15$ alone can reionize the IGM \citep{finkelstein2012, bouwens2012, duncan2015, robertson2015}, although \citet{naidu2019} assume $f_{\rm esc} \propto$ the star formation rate surface density and infer that high stellar mass (${\rm M_*} \gsim 10^8\msun$) galaxies dominate the reionization budget \citep[see also][]{sharma2016}. The bulk of the observational results are in agreement with theoretical results that converge on stars in low-mass halos ($\mh \lsim 10^{9.5} \Msun$ and $\muv \gsim -17$) providing the bulk of \HI ionizing photons at $z \gsim 7$ \citep[e.g.][]{choudhury2007, salvaterra2011, yajima2011, wise2014, paardekooper2015, liu2016, dayal2017reio}. A key caveat in the results, however, is that the redshift-dependent reionization contribution from star formation in galaxies of different masses crucially depends on the strength of UVB feedback, the trend of $f_{\rm esc}$ with mass and redshift and the evolution of the clumping factor \citep[for details see Sec. 7,][]{dayal2018}.  

% choudhury2007, razoumov2010, salvaterra2011, raicevic2011, finlator2011, yajima2011, paardekooper2011, wise2014, paardekooper2015, liu2016,qin2017, dayal2017reio

In addition, the contribution of Active Galactic Nuclei (AGN) to reionization and its dependence on redshift and on the host galaxy stellar mass still remain key open questions. A number of works show AGN can only have a minor reionization contribution \citep{yoshiura2017, onoue2017, hassan2018}. Contrary to these studies, a number of results show that radiation from AGN/quasars might contribute significantly to reionization \citep{volonteri2009, madau2015, mitra2015, mitra2018, giallongo2018, finkelstein2019}, especially at $z \gsim 8$ if ionizations by secondary electrons are accounted for, with stars taking over as the dominant reionization sources at $z \lsim 6$ \citep{volonteri2009}. The question of the contribution of AGN to reionization has witnessed a resurgence after recent claims of extremely high number densities of faint AGN measured by \citet{giallongo2015, giallongo2019} at $z \gsim 4$. While other direct searches for high-redshift AGN have found lower number densities \citep{weigel2015, mcgreer2018}, the integrated \HI ionizing emissivities can be significantly affected by the inhomogeneous selection and analysis of the data and by the adopted (double) power law fits to the AGN luminosity function at different redshifts \citep{Kulkarni2019}. Yet, if the high comoving emissivity claimed by \citet{giallongo2015} persists up to  $z \simeq 10$, then AGN alone could drive reionization with little/no contribution from starlight \citep{madau2015}. A similar scenario, where more than 50\% of the ionizing photons are emitted by rare and bright sources, such as quasars, has been proposed by \citet{chardin2015, chardin2017} as a possible explanation of the large fluctuations in the Ly$\alpha$ effective optical depth on scales of 50 $h^{-1}$ cMpc measured at the end stages of reionization ($4 < z < 6$) by \citet{becker2015}. These AGN-dominated or AGN-assisted models, however, are found to reionize helium (\HeII) too early \citep{puchwein2019} and result in an IGM temperature evolution that is inconsistent with the observational constraints \citep{becker2011}.

In this work, we use a semi-analytic model ({\it Delphi}) that has been shown to reproduce all key observables for galaxies and AGN at $z \gsim 5$ to revisit the AGN contribution to reionization, specially as a function of the host galaxy stellar mass. The key strengths of this model lie in that: (i) it is seeded with two types of black hole seeds (stellar and direct collapse); (ii) the black hole accretion rate is primarily regulated by the host halo mass; (iii) it uses a minimal set of free parameters for star formation and black holes and their associated feedback.

%However, in addition to the fact that later works have not been able to corroborate these high AGN number densities, AGN-driven reionization results in a slower evolution of the HeII Ly$\alpha$ forest as compared to observations and is therefore ruled out at the $2-\sigma$ level; the data therefore favours a model where galaxies with $f_{esc} \sim 10\%$ dominate in the initial reionization stages with AGN being relevant only at $z \lsim 6$ \citep{mitra2018}. 

The cosmological parameters used in this work correspond to $\Omega_{\rm m },\Omega_{\Lambda}, \Omega_{\rm b}, h, n_s, \sigma_8 = 0.3089,0.6911,0.049, 0.67, 0.96, 0.81$ \citep{planck2015}. We quote all quantities in comoving units unless stated otherwise and express all magnitudes in the standard AB system \citep{oke-gunn1983}.

The paper is organized as follows. In Section \ref{theo}, we detail our code for the galaxy-BH (co)-evolution, our calculation of $f_{\rm esc}$ and the progress of reionization. The results of the fiducial and of alternative models are presented in Sections \ref{results} and \ref{alt_models}. Finally, we discuss our results and present our main conclusions in Section \ref{sec:conclusions}.

%*******************************************************************
\begin{table*}
\begin{center}
\caption{Free parameters, their symbols and values used for the {\it fiducial} model \citep[{\it ins1} in][]{dayal2019}. As noted, using these parameter values our model reproduces all key observables for galaxies and AGN at $z \gsim 5$ (including their UV luminosity functions, stellar mass/black hole mass densities, star formation rate densities, the stellar/black hole mass function) as well as the key reionization observables (the integrated electron scattering optical depth and the redshift evolution of the ionizing photon emissivity). Simultaneously fitting the optical depth and the emissivity constraints, we obtain $f_0=0.02 \, (0.0185)$ and $\beta=2.8 \, (2.8)$ if we consider the ionizing photons provided by star formation (star formation and AGN).}
\begin{tabular}{|c|c|c|}
\hline
Parameter & Symbol & value \\
\hline
Maximum star formation efficiency &  $f_*$ & $0.02$ \\
Fraction of SNII energy coupling to gas & $f_{\rm w}$ &   $0.1$ \\
Radiative efficiency of black hole accretion & $\epsilon_{\rm r}$ & $0.1$ \\
Fraction of AGN energy coupling to gas &  $f_{\rm bh}^{\rm w}$ & $0.003$ \\
Fraction of gas mass AGN can accrete & $f_{\rm bh}^{\rm ac}$ &  $5.5 \times 10^{-4}$\\
Fraction of Eddington rate for BH accretion & $f_{\rm Edd} (\mh < {\rm M_h^{crit}})$ & $7.5 \times 10^{-5}$ \\
Fraction of Eddington rate for BH accretion & $f_{\rm Edd} (\mh \geq {\rm M_h^{crit}})$ & $1$ \\
LW BG threshold for DCBH formation & $\alpha$ &  $30$ \\
Escape fraction of \HI ionizing photons from star formation & $\fescsf$ &  $f_0 [(1+z)/7]^\beta$. \\
Escape fraction of \HI ionizing photons from AGN & $f_{\rm esc}^{\rm bh}$ & \citet{2014ApJ...786..104U} \\
Stellar population synthesis model & - &  {\it Starburst99} \\
Reionization (UVB) feedback & - & No \\
\hline
\end{tabular}
\label{table1}
\end{center}
\end{table*}

%*******************************************************************
% *************************************************************
\section{Theoretical model}
\label{theo}
% *************************************************************
We start by introducing the galaxy formation model in Sec. \ref{delphi} before discussing the escape fraction of ionizing radiation from galaxies and AGN in the fiducial model in Sec. \ref{sec_fesc}. These are used to calculate the reionization history and electron scattering optical depth in Sec. \ref{sec_reio}. Our fiducial model parameters are described in Table \ref{table1}.

% ************************
\subsection{Galaxy formation at high-$z$}
\label{delphi}
% ************************

In this work, we use the semi-analytic code \textit{Delphi} (\textbf{D}ark matter and the \textbf{e}mergence of ga\textbf{l}axies in the e\textbf{p}oc\textbf{h} of re\textbf{i}onization) that aims at simulating the assembly of the dark matter, baryonic and black hole components of high-redshift ($z \gsim 5$) galaxies \citep{dayal2014a, dayal2019}. In brief, starting at $z=4$ we build analytic merger trees up to $z=20$, in time-steps of 20 Myrs, for 550 haloes equally separated in log space between $10^8$-$10^{13.5}\ \Msun$. Each halo is assigned a number density according to the Sheth-Tormen halo mass function (HMF) which is propagated throughout its merger tree; the resulting HMFs have been confirmed to be in accord with the Sheth-Tormen HMF at all $z \sim 5-20$. 

The very first progenitors of any galaxy are assigned an initial gas mass as per the cosmological baryon-to-dark matter ratio such that ${\rm M_{gi}} = (\Omega_{\rm b}/\Omega_{\rm m}) \mh$, where $\mh$ is the halo mass. The effective star formation efficiency, $f_*^{\rm eff}$, for any halo is calculated as the minimum between the efficiency that produces enough type II supernova (SNII) energy to eject the rest of the gas, $f_*^{\rm ej}$, and an upper maximum threshold, $f_*$, so that $f_*^{\rm eff} = {\rm min}[f_*^{\rm ej}, f_*]$ where a fraction $f_{\rm w}$ of the SNII energy can couple to the gas. The gas mass left after including the effects of star formation and supernova feedback is then given by:
\begin{equation}
{\rm M_*^{gf}}(z) = [{\rm M_{gi}}(z)-{\rm M_*}(z)] \bigg(1-\frac{f_*^{\rm eff}}{f_*^{\rm ej}}\bigg). 
\label{eq:gasmass}
\end{equation}
\noindent
Our model also includes two types of black hole seeds that can be assigned to the first progenitors of any halo. These include {\it (i)} massive direct-collapse black hole (DCBH) seeds with masses between $\mbh =10^{3-4}\msun$ and, {\it (ii)} PopIII stellar black hole seeds of $150 \msun$ masses. As detailed in \citet{dayal2017dcbh}, we calculate the strength of the Lyman-Werner (LW) background irradiating each such starting halo. Halos with a LW background strength $J_{\rm LW} > J_{\rm crit} = \alpha J_{\rm 21}$ (where $J_{\rm 21} = 10^{-21} \, \mathrm{erg s^{-1} Hz^{-1} cm^{-2} sr^{-1}}$ and $\alpha$ is a free parameter) are assigned DCBH seeds while halos not meeting this criterion are assigned the lighter PopIII seeds. We note that, given that the number densities of DCBH seeds are $\sim -2 \, (-3.8)$ orders of magnitude below that of stellar seeds for $\alpha = 30\, (300)$, the exact value of $\alpha$ (as well as the DCBH seed mass) have no sensible bearing on our results, since we only consider models that reproduce the AGN luminosity function. In this paper we do not aim at investigating which type of black hole seed can contribute most to reionization, but how a population of AGN reproducing available observational constraints can contribute to reionization.

Once seeded, the black holes (as the baryonic and dark matter components) grow in mass through mergers and accretion in successive time-steps. %At any given time-step, the star formation efficiency is calculated as being the minimum between the SNII energy required to unbind the rest of the gas and a effective threshold value $f_*^{eff}$. 
A fraction of the gas mass left after star formation and SNII ejection (see Eqn. \ref{eq:gasmass}) can be accreted onto the black hole. This accretion rate depends on both the host halo mass and redshift through a critical halo mass \citep{bower2017}:
\begin{equation}
{\rm M_{h}^{crit}}(z) = 10^{11.25} \msun [\Omega_{\rm m}(1+z)^3 + \Omega_\lambda]^{0.125}, 
\label{charmh}
\end{equation}
\noindent
such that the mass accreted by the black hole (of mass $\mbh$) at any given time-step is: 
\begin{equation}
{\rm M_{bh}^{ac}}(z) = \min\left[f_{\rm Edd}{\rm M_{Edd}}(z),\ (1-\epsilon_{\rm r})f_{\rm bh}^{\rm ac}{\rm M_*^{gf}}(z) \right],
\label{accchannel}
\end{equation}
\noindent
where ${\rm M_{Edd}}(z) = (1-\epsilon_r) [4 \pi G \mbh (z) m_p][\sigma_T \epsilon_r c]^{-1} \Delta t$ is the total mass that can be accreted in a time-step assuming Eddington luminosity. Here, $G$ is the gravitational constant, $m_p$ is the proton mass, $\sigma_T$ is the Thomson scattering optical depth, $\epsilon_r$ is the BH radiative efficiency, $c$ is the speed of light and $\Delta t = 20 {\rm Myr}$ is the merger tree time-step. Further, the value of $f_{\rm Edd}$ is assigned based on the critical halo mass (Eqn. \ref{charmh}) as detailed in Table \ref{table1} and 
$f_{\rm bh}^{\rm ac}$ represents a fixed fraction of the total gas mass present in the host galaxy that can be accreted by the black hole. A fixed fraction $f_{\rm bh}^{\rm w}$ of the total energy emitted by the accreting black hole is allowed to couple to the gas content. The values used for each of these parameters in our fiducial model are detailed in Table \ref{table1}. Finally, reionization feedback is included by suppressing the gas content, and hence star formation and black hole accretion, of halos with a virial velocity $\vvir \lsim 40\, {\rm km\, s^{-1}}$ at all redshifts, as detailed in Sec. \ref{sec_reio}.

%\begin{equation}
%f_{Edd} = \begin{cases} 
 %     7.5 \times 10^{-5} & M_h(z) < M_{bh}^{crit}(z) \\
 %     1 & M_h(z) \geq M_{bh}^{crit}(z) \\
 %  \end{cases}
 %  \label{eq_fedd}
%\end{equation}

In the interest of simplicity, every newly formed stellar population is assumed to follow a Salpeter initial mass function \citep[IMF;][]{salpeter1955} with masses in the range $0.1-100\Msun$, with a metallicity $Z = 0.05 Z_\odot$ and an age of $2\,\myr$; a lower (higher) metallicity or a younger (older) stellar population across all galaxies would scale up (down) the UV luminosity function which could be accommodated by varying the free-parameters for star formation ($f_*^{\rm eff}$ and $f_{\rm w}$). Under these assumptions, the Starburst99 ({\it SB99}) stellar population synthesis (SPS) model yields the time-evolution of the star-formation powered production rate of \HI ionizing photons ($\dot n_{\rm int}^{\rm sf}$) and the UV luminosity ($\luv$) to be:
\begin{equation}
\dot n_{\rm int}^{\rm sf}(t) = 10^{46.6255} - 3.92 \, {\rm log10} \bigg(\frac{t}{2\ \myr} \bigg) + 0.7 \, [{\rm s^{-1}}],
\end{equation}
and
\begin{equation}
L_{\rm UV}(t) = 10^{33.077} - 1.33 \, {\rm log10} \bigg(\frac{t}{2\ \myr} \bigg) + 0.462 \, [{\rm erg\, s^{-1} \, \AA^{-1}}].
\label{uv_sb99}
\end{equation}

Inspired by the Shakura-Sunyaev solution \citep{Shakura1973}, AGN are assigned a spectral energy distribution (SED) that depends on the key black hole physical parameters, namely the black hole mass and Eddington ratio \citep{2017ApJ...849..155V}. We follow here a variant based on the physical models developed by \cite{2012MNRAS.420.1848D}. Specifically, we calculate the energy of the peak of the SED as described in \cite{2016ApJ...833..266T},  but adopt the default functional form of the spectrum used in Cloudy \citep{2013RMxAA..49..137F}. 

Once an AGN is assigned a luminosity and a SED, the UV luminosity is calculated as detailed in \citet{dayal2019}. Further, we integrate above 13.6~eV to obtain the H\,\textsc{i}  ionizing luminosity and mean energy of ionizing photons (see Fig. \ref{fig_app} in the Appendix). For AGN, this  provides an upper limit, as photons above 24.59~eV and 54.4~eV can ionize He\,\textsc{i} and He\,\textsc{ii}. We further include a correction for secondary ionizations from the hard AGN photons, by taking the upper limit to their contribution, i.e., assuming fully neutral hydrogen and that 39\% of their energy goes into secondary ionizations \citep{Shull1985,2017ApJ...840...39M}.

%To estimate the statistical contribution of galaxies of different mass, and of the AGN they host we weight their contribution by their number densities assigned as explained above. 
%In the following we  use as a reference the galaxy mass function by \cite{2016ApJ...825....5S}, extrapolated down to $10^5 \msun$ and up to $3\times10^{12} \msun$; see \cite{2013ApJ...770...57B},  \cite{2014ARA&A..52..415M}, and \cite{2016ARA&A..54..761S} for differences and uncertainties among various determinations. 
% ************************
\subsection{The escape fraction of \HI ionizing photons}
\label{sec_fesc}
% ************************
In what follows, we discuss our calculations of $f_{\rm esc}$ for both AGN and stellar radiation from galaxies. In addition to the fiducial model, we study five combinations of $f_{\rm esc}$ from star formation and AGN in order to explore the available parameter space and its impact on our results as detailed in Sec. \ref{alt_models}.

% *********************************
\subsubsection{The escape fraction for AGN ($f_{\rm esc}^{\rm bh}$)}
\label{fesc_agn}

For the ionizing radiation emitted from the AGN, we consider four different models. We start by taking an approach similar to \cite{2017MNRAS.465.1915R} for the
fiducial model. Essentially, we assume that the unobscured fraction, i.e., the fraction of AGN with column density $<10^{22}\, {\rm cm}^{-2}$ is a proxy for the escape fraction, $f_{\rm esc}^{\rm bh}$. The argument is that by applying a column-density dependent correction to the X-ray LF, one recovers the UV luminosity function. As in \citet{dayal2019}, we adopt the luminosity-dependent formalism of \cite{2014ApJ...786..104U}, taking as unobscured fraction $f_{\rm unabs}\equiv f_{\rm logNH<22}$, which varies from $\simeq$ 10\% for faint AGN ($ L_{\rm 2-10keV}<10^{43}$ erg s$^{-1}$) to $\simeq$ 67\% for bright AGN ($L_{\rm 2-10keV}>10^{46}$ erg s$^{-1}$). The unobscured fraction can be written as:
\begin{equation}
f_{\rm unabs}=\frac{1-\psi}{1+\psi},
\end{equation} 
where $\psi=\psi_{\rm z}-0.24(L_{\rm x}-43.75)$, $\psi_{\rm z}=0.43[1+\min(z,2)]^{0.48}$ and $L_{\rm x}$ is the log of the intrinsic 2--10 keV X-ray luminosity in erg~s$^{-1}$; given our model is for $z \gsim 5$, this implies $\psi_{\rm z}=0.73$. We do not extrapolate the evolution beyond $z=2$, the range for which the dependence has been studied using data. As in \cite{2017MNRAS.465.1915R}, we assume that unobscured quasars have $\fesc=1$ and zero otherwise \citep[see their section 4.1 for a discussion and alternative models and][for a discussion on the redshift evolution of the obscured fraction]{2017ApJ...849..155V}.  

Secondly, \cite{merloni2014} find that X--ray and optical obscuration are not necessarily the same for AGN, although the trend of optically obscured AGN with luminosity is consistent with the scaling we adopt. Our second model for  $\fesc^{\rm bh}$ considers the fraction of optically unobscured AGN as a function of luminosity from \cite{merloni2014}, where this fraction is found to be independent of redshift.  It takes the functional form:
\begin{equation}
\fesc^{\rm bh}=1-0.56+\frac{1}{\pi}\arctan\left( \frac{43.89 -\log L_{\rm x}}{0.46} \right),
\label{fesc_merloni}
\end{equation} 
where $\log L_{\rm x}$ is the logarithm of the intrinsic 2--10 keV X-ray luminosity in erg~s$^{-1}$. 

Thirdly, we can maximize the contribution of AGN to reionization by assuming $\fesc^{\rm bh}=1$, although \cite{Micheva2017} find that even for unobscured AGN $\fesc^{\rm bh}$ is not necessarily unity. 

Finally, we explore a model wherein we use the same (redshift-dependent) escape fraction for the ionizing radiation from both star formation and AGN. The results from these last three cases are discussed in detail in Sec. \ref{alt_models}.

% *********************************
\subsubsection{The escape fraction for star formation ($\fescsf$)}
\label{fesc_sf}
% *********************************
Both the value of the escape fraction of \HI ionizing radiation emitted from the stellar population ($\fescsf$) as well as its trend with the galaxy mass or even redshift remain extremely poorly understood \citep[Sec. 7.1,][]{dayal2018}. We study four cases for $\fescsf$ in this work: firstly, in our fiducial model, we use an escape fraction that scales down with decreasing redshift as $\fescsf = f_0 [(1+z)/7]^\beta$ where $\beta>1$ and $f_0$ is a constant at a given redshift. This is in accord with a number of studies \citep{robertson2015, dayal2017reio, puchwein2019} that have shown that simultaneously reproducing the values of electron scattering optical depth ($\tau_{\rm es}$) and the redshift evolution of the emissivity require such a decrease in the global value of the escape fraction of ionizing photons from star formation. The values of $f_0$ and $\beta$ required to simultaneously fit the above-noted data-sets (with and without AGN contribution) are shown in Table \ref{table1}. 

Secondly, whilst maintaining the same functional form, we find the values of the two coefficients ($f_0$ and $\beta$) required to fit the optical depth and emissivity constraints using the same escape fraction from AGN and star formation. 

Thirdly, following recent results \citep[e.g.][]{borthakur2014, naidu2019}, we use a model wherein the escape fraction for star formation scales positively with the stellar mass. In this case, for galaxies that have black holes, we assume $f_{esc}^{sf}=\fescbh$ using the fiducial model for $\fescbh$; $f_{\rm esc}^{sf}=0$ for galaxies without a black hole. This accounts for the possibility that AGN feedback enhances the effect of SN feedback in carving ``holes" in the interstellar medium, facilitating the escape of ionizing radiation. This is a very optimistic assumption, as dedicated simulations show that AGN struggle to shine and amplify the escape fraction in low-mass galaxies \citep{Trebitsch2018}. 

Fourthly, we explore a model with a constant $\fescsf = 0.035$. Although a constant escape fraction for stellar radiation from all galaxies can reproduce the $\tau_{\rm es}$ value, it over-shoots the vale of the observed emissivity \citep[see e.g. Fig. 3,][]{dayal2017reio}. 

Finally, we explore a model wherein $\fescsf$ increases with decreasing stellar mass, as has been shown by a number of theoretical works \citep[e.g.][]{yajima2011, wise2014, paardekooper2015}. Essentially, we assume $\fescsf$ scales with the ejected gas fraction such that $\fescsf = f_0 (f_*^{\rm eff}/f_*^{\rm ej})$. This naturally results in a high $\fescsf$ value for low mass galaxies where $f_*^{\rm eff}=f_*^{\rm ej}$; $\fescsf$ drops with increasing mass where $f_*^{\rm eff} \sim f_* < f_*^{\rm ej}$. The results from these last four cases are discussed in detail in Sec. \ref{alt_models}.

We clarify that while we assume the same $\fescsf$ value for each galaxy, in principle, this should be thought of as an ensemble average that depends on, and evolves with, the underlying galaxy properties, such as mass or star formation or a combination of both. 

\subsection{Modelling reionization}
\label{sec_reio}
% *************************************************************
The reionization history, expressed through the evolution of the volume filling fraction ($Q_{\rm II}$) for ionized hydrogen (\HII), can be written as \citep{shapiro-giroux1987, madau1999}:
\begin{equation}
\label{filfracz}
\frac{dQ_{\rm II}}{dz} = \frac{dn_{\rm ion}} {dz}\frac{1}{n_{\rm H}} - \frac{Q_{\rm II}}{t_{\rm rec}} \frac{dt}{dz},
\end{equation}
where the first term on the right hand side is the source term while the second term accounts for the decrease in $Q_{\rm II}$ due to recombinations. Here, $dn_{\rm ion}/dz=\dot n_{ion}$ represents the hydrogen ionizing photon rate density contributing to reionization. Further, $n_{\rm H}$ is the comoving hydrogen number density and $t_{\rm rec}$ is the recombination timescale that can be expressed as \citep[e.g.][]{madau1999}:
\begin{equation}
t_{\rm rec} = \frac{1}{\chi\, n_{\rm H} \, (1+z)^3 \alpha_{\rm B} \, C}.
\end{equation}
Here $\alpha_{\rm B}$ is the hydrogen case-B recombination coefficient, $\chi = 1.08$ accounts for the excess free electrons arising from singly ionized helium and $C$ is the IGM clumping factor. We use a value of $C$ that evolves with redshift as
\begin{equation}
C = \frac{<n_{\rm HII}^2>}{<n_{\rm HII}>^2} = 1+ 43 \, z^{-1.71}.
\end{equation}
using the results of \citet{pawlik2009} and \citet{haardt-madau2012} who show that the UVB generated by reionization can act as an effective pressure term, reducing the clumping factor.

\begin{figure*}
\begin{center}
\center{\includegraphics[scale=1.05]{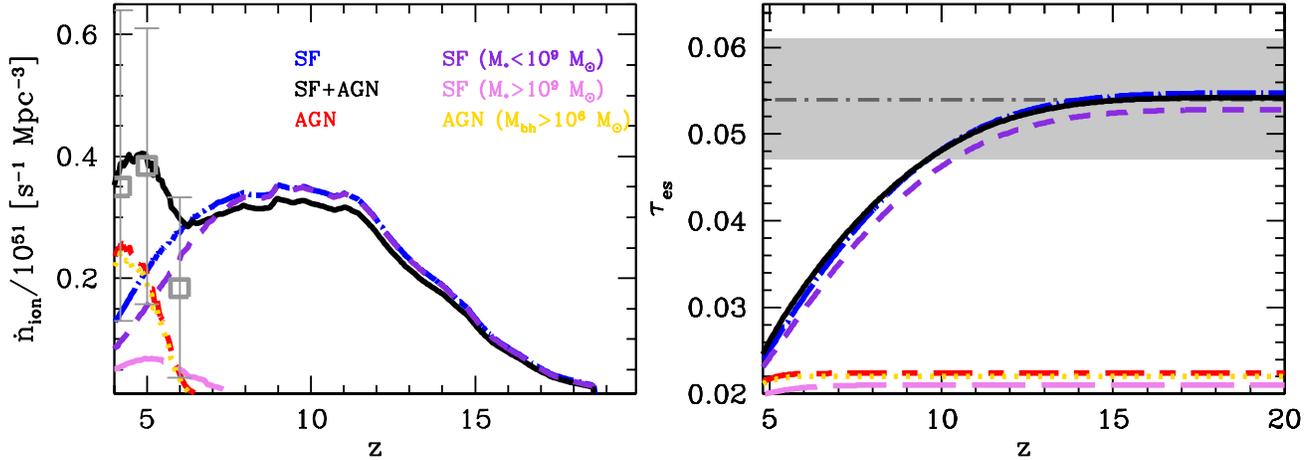}} % nionfnz_taufnz_qiifnz_sb99_gal_agn_galgl9_bhg9.pdf
\caption{Redshift evolution of the \HI ionizing photon emissivity ({\it left panel}) and the CMB electron scattering optical depth ($\tau_{\rm es}$) as a function of redshift ({\it right panel}) for the {\it fiducial} model. In the left panel, the open squares show observational results (and associated error bars) calculated following the approach of \citet{kfg2012}. In the right panel, the dot-dashed horizontal line shows the central value for $\tau_{\rm es}$ inferred by the latest Planck results \citep{planck2018} with the gray striped region showing the $1-\sigma$ errors. Over-plotted are the escaping emissivities ({\it left panel}) and the optical depths ({\it right panel}) contributed by: star formation only (SF; dot-long-dashed line), AGN+star formation (solid line), and AGN only (short-long-dashed line) using the $\fescsf$ and $\fesc^{\rm bh}$ values for the fiducial model reported in Table \ref{table1}; note that $\fescsf$ is lower in the AGN+SF case ($f_0=0.0185$) as compared to the SF only case ($f_0=0.02$). We deconstruct the contribution from star formation in galaxies into those with stellar masses $\mstar \lsim 10^9 \msun$ (short-dashed line) and $\mstar \gsim 10^9 \msun$ (long-dashed line) and show the contribution of black holes of masses $\gsim 10^6 \msun$ using the dotted line, as marked.}
\label{fig_emm}
\end{center}
\end{figure*}

\begin{figure*}
\begin{center}
\center{\includegraphics[scale=1.04]{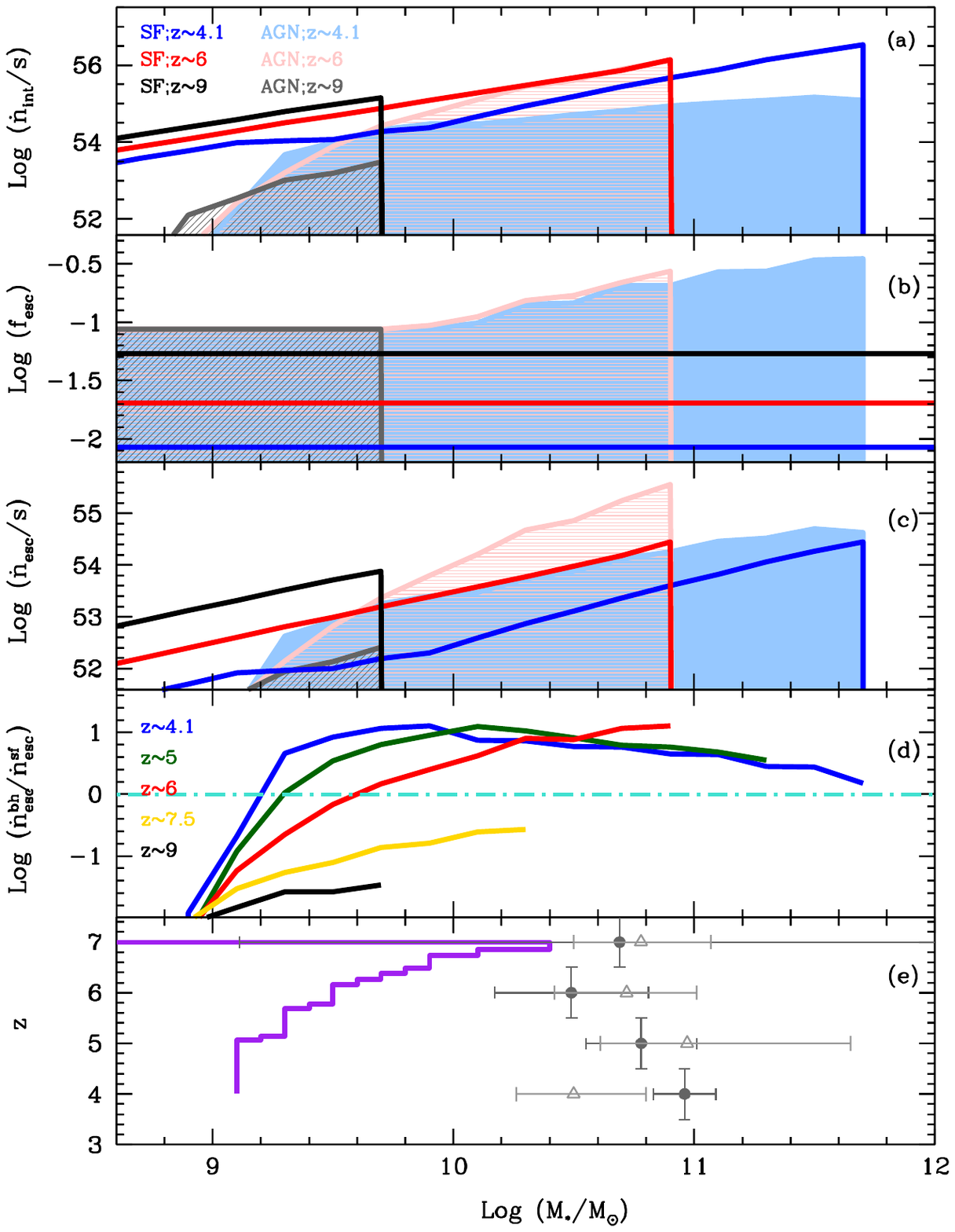}} 
\caption{As a function of stellar mass, the panels (top to bottom) show the results for star formation (solid lines) and AGN (light shaded regions) for the {\it fiducial} model for: {\it (a):} the {\it intrinsic} \HI ionizing photon rate; {\it (b):} the escape fraction of \HI ionizing photons; {\it (c):} the {\it escaping} \HI ionizing photon rate;  {\it (d):} the ratio between the {\it escaping} \HI ionizing photon rate for AGN and star formation with the horizontal line showing a ratio of unity; and {\it (e):} the transition stellar mass at which AGN start dominating the escaping ionizing photon production rate. In this panel, the solid circles and empty triangles show the knee value of the stellar mass function (and the associated error bars) observationally inferred by \citet{grazian2015} and \citet{song2016}, respectively. Finally, the different colours in panels (a)-(c) are for the redshifts marked in panel (a) while the different lines in panel (d) are for the redshifts marked in that panel. }
\label{fig_ife}
\end{center}
\end{figure*}

While reionization is driven by the hydrogen ionizing photons produced by stars in early galaxies, the UVB built up during reionization suppresses the baryonic content of galaxies by photo-heating/evaporating gas at their outskirts \citep{klypin1999, moore1999, somerville2002}, suppressing further star formation and slowing down the reionization process. In order to account for the effect of UVB feedback on $\dot n_{\rm ion}$, we assume total photo-evaporation of gas from halos with a virial velocity below $\vvir=40 \, {\rm km\, s^{-1}}$ embedded in ionized regions at any $z$. In this ``maximal external feedback" scenario, halos below $\vvir$ in ionized regions neither form stars nor contribute any gas in mergers. 

The globally averaged $\dot n_{\rm ion}$ can then be expressed as:
\begin{equation}
\label{eq_nion}
  \dot n_{\rm ion}(z) = \dot n_{\rm esc}^{\rm sf}(z) + \dot n_{\rm esc}^{\rm bh}(z)
\end{equation}
\noindent
where
\begin{eqnarray}
\label{nesc}
\dot n_{\rm esc}^{\rm sf}(z)&=&\fescsf [Q_{\rm II}(z) \dot n_{\rm int,II}^{\rm sf}(z)  + Q_{\rm I}(z) \dot n_{\rm int,I}^{\rm sf}(z) ], \\
\dot n_{\rm esc}^{\rm bh}(z)&=&f_{\rm esc}^{\rm bh}[Q_{\rm II}(z) \dot n_{\rm int,II}^{\rm bh}(z) + Q_{\rm I}(z) \dot n_{\rm int,I}^{\rm bh}(z)],
\end{eqnarray}
\noindent
where $Q_{\rm I}(z) = 1-Q_{\rm II}(z)$. Further, $\dot n_{\rm int,II}^{\rm sf}$ ($\dot n_{\rm int,II}^{\rm bh}$) and $\dot n_{\rm int,I}^{\rm sf}$ ($\dot n_{\rm int,I}^{\rm bh}$) account for the intrinsic hydrogen ionizing photon 
production rate density from star formation (black hole accretion) in case of full UV-suppression of the gas mass and no UV suppression, respectively. 
The term $\dot n_{\rm esc}^{\rm sf}$ ($\dot n_{\rm esc}^{\rm bh}$) weights these two contributions over the volume filling fraction of ionized and neutral regions -  i.e. while $\dot n_{\rm int,I}$ represents 
the contribution from all sources, stars and black holes in halos with $\vvir < 40 \, {\rm km} {\rm s}^{-1}$ do not contribute to $\dot n_{\rm int, II}$. At the beginning of the reionization process, the volume filled by ionized hydrogen is very small ($Q_{\rm II}<<1$) and most galaxies are not affected by UVB-feedback, so that $\dot n_{\rm ion}(z) \approx \dot n_{\rm int,I}^{\rm sf}(z) \fescsf \, +\dot n_{\rm int,I}^{\rm bh}(z) f_{\rm esc}^{\rm bh}$. As $Q_{\rm II}$ increases and reaches a value $\simeq 1$, all galaxies in halos with circular velocity less than $\vvir = 40\, {\rm km}\, {\rm s}^{-1}$ are feedback-suppressed, so that $\dot n_{\rm ion}(z) \approx \dot n_{\rm int,II}^{\rm sf}(z) \fescsf \, +\dot n_{\rm int,II}^{\rm bh}(z) f_{\rm esc}^{\rm bh}$.

% *******************************************
\section{Results}
\label{results}
% *******************************************

Given that $\dot n_{\rm ion}(z)$ is an output of the model, $t_{\rm rec}$ is calculated as a function of $z$ and $f_{\rm esc}^{\rm bh}$ is obtained from the AGN obscuration fraction, $\fescsf$  is the only free parameter in our reionization calculations. As explained above, in the {\it fiducial} model, $\fescsf$ is composed of two free parameters ($f_0$ and $\beta$) that are fit 
by jointly reproducing the observed values of $\tau_{\rm es}$ and the emissivity as discussed in Sec. \ref{obs_reio} that follows. We use this $\fescsf$ value to study the AGN contribution to reionization in Sec. \ref{agn_reio}. In order to test the robustness of our results to assumptions, we also explore alternative models for the escape fraction from AGN and star formation and the impact of different stellar population synthesis models in Sec. \ref{alt_models}.

% *******************************************
\subsection{The electron scattering optical depth and the ionizing photon emissivity}
\label{obs_reio}
% *******************************************
We start by discussing the redshift evolution of the ionizing photon emissivity (Eqn. \ref{eq_nion}) from the fiducial model shown in the left panel of Fig. \ref{fig_emm}. 
For star formation, the ``escaping" emissivity includes the effect of $\fescsf$ that decreases with redshift as $\propto [(1+z)/7]^{2.8}$. As a result, whilst increasing from $z \sim 19$ to $z\sim 8$ the emissivity from stellar sources in galaxies thereafter shows a drop at lower redshifts. Low-mass ($\mstar \lsim 10^9 \msun$) galaxies dominate the stellar emissivity at all redshifts and the total (star formation+AGN) emissivity down to $z \sim 5$; although sub-dominant, the importance of stars in massive ($\mstar \gsim 10^9 \msun$) galaxies increases with decreasing redshift and they contribute as much as 40\% ($\sim 15\%$) to the stellar (total) emissivity at $z \sim 4$. 

On the other hand, driven by the growth of black holes and the constancy of  $f_{\rm esc}^{\rm bh}$ with redshift, the AGN emissivity shows a steep (six-fold) increase in the 370 Myrs between $z \sim 6$ and $4$. A turning point is reached at $z \sim 5$ where AGN and star formation contribute equally to the total emissivity, with the AGN contribution (dominated by $M_{\rm bh} \gsim 10^6 \msun$ black holes in $\mstar \gsim 10^{9}\msun$ galaxies) overtaking that from star formation at lower-$z$. Indeed, the AGN emissivity is almost twice of that provided by stars by $z \sim 4$ leading to an increase in the {\it total} value. 

\begin{figure*}
\begin{center}
\center{\includegraphics[scale=0.75]{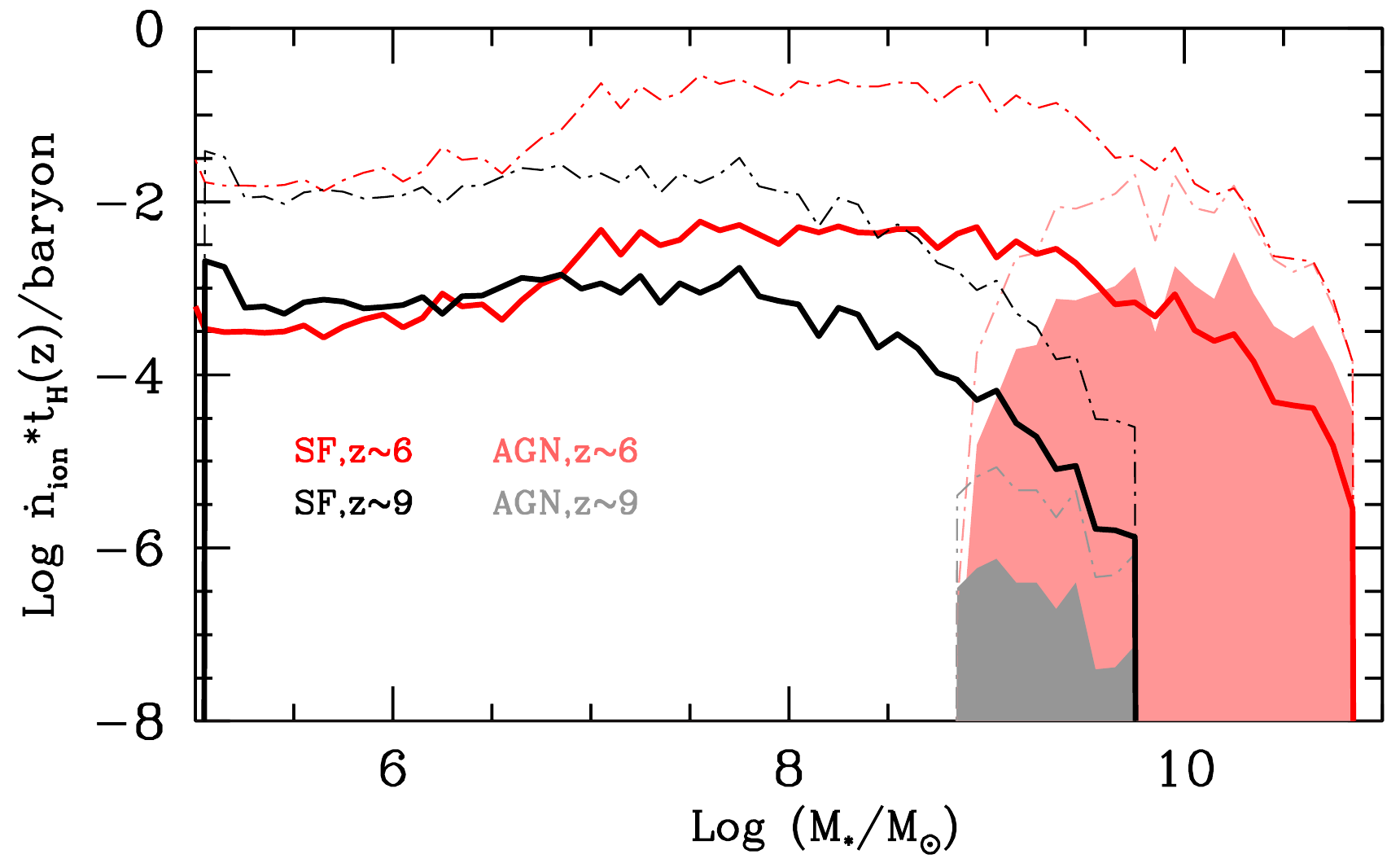}} 
\caption{The ionizing photon per baryon value as a function of stellar mass for the {\it fiducial} model for star formation and AGN at $z \sim 6$ and $9$, as marked. The dot-dashed and solid lines show the {\it intrinsic} and {\it escaping} \HI ionizing photon rates, respectively. }
\label{fig_bar}
\end{center}
\end{figure*}

To summarise, while the trend of the total emissivity is driven by star formation in low-mass galaxies down to $z = 5$, AGN take over as the dominant contributors at lower redshifts. This result is in agreement with synthesis models for the UVB \citep{fg2008,haardt-madau2012} as shown in the same figure. 

The above trends can also be used to interpret the latest results on the integrated electron scattering optical depth \citep[$\tau_{\rm es}=0.054\pm0.007$;][]{planck2018}, shown in the right panel of Fig. \ref{fig_emm}. We start by noting that fitting to this data requires $\fescsf = 0.02 [(1+z)/7]^{2.8}$ if stars in galaxies are considered to be the only reionization sources; as shown in Table \ref{table2} considering the contribution of both stars and AGN leads to a marginal decrease in the co-efficient of $\fescsf$ to 0.0185 whilst leaving the redshift-relation unchanged. Stellar radiation in low-mass ($\mstar \lsim 10^9 \msun$) galaxies dominate the contribution to $\tau_{\rm es}$ for most of reionization history. AGN only start making a noticeable contribution at $z \lsim 5$, where they can generate an optical depth of $\tau_{\rm es} \sim 0.22$, comparable to stars, which generate a total value of $\tau_{\rm es} \sim 0.24$. Stellar radiation from high-mass ($\mstar \gsim 10^9 \msun$) galaxies has a sub-dominant contribution to $\tau_{es}$ at all redshifts. 

% *******************************************
\subsection{AGN contribution to reionization as a function of stellar mass}
\label{agn_reio}
% *******************************************
To understand the AGN contribution to reionization in the {\it fiducial} model, we start by looking at the (intrinsic) production rate of \HI ionizing photons as a function of $\mstar$ for $z \sim 4-9$ (panel a; Fig. \ref{fig_ife}). As expected, $\dot n_{\rm int}^{\rm sf}$ scales with $\mstar$ since higher mass galaxies typically have larger associated star formation rates. Further, given their larger gas and black hole masses, $\dot n_{\rm int}^{\rm bh}$ too scales with $\mstar$. As seen, stars dominate the {\it intrinsic} \HI ionizing radiation production rate for all stellar masses at $z \gsim 7$. However, moving to lower redshifts, black holes can contribute as much as stars in galaxies with $\mstar \sim 10^{10.2-10.9}\msun$ at $z \sim 6$. This mass range decreases to $\mstar \sim 10^{9.6-10}\msun$ at $z \sim 4$ where intermediate-mass galaxies host black holes that can accrete at the Eddington rate. 

The second factor that needs to be considered is the escape fraction of ionizing photons which is shown in panel (b) of the same figure. As noted above, $\fescsf$ is independent of galaxy properties and decreases with decreasing $z$, going from a value of about $5.4\%$ at $z \sim 9$ to $0.77\%$ at $z\sim 4$. 

However, $f_{\rm esc}^{\rm bh}$ scales with $\mstar$, and this is the result of the dependence of the unabsorbed AGN fraction with luminosity: at higher AGN luminosity a higher fraction of AGN are unabsorbed. Quantitatively, while $f_{\rm esc}^{\rm bh} \sim 10\%$ for $\mstar \lsim 10^{9.7}\msun$, it can have a value as high as $30\%$ for $\mstar \gsim 10^{10.9}\msun$ at $z \sim 6-9$.

We can now combine the intrinsic production rate of \HI ionizing photons and the escape fraction to look at the rate of ``escaping" ionizing radiation for star formation and AGN in panel (c) of Fig. \ref{fig_ife}. As expected, $\dot n_{\rm esc}^{\rm sf} \propto \mstar$ and $\dot n_{\rm esc}^{\rm sf}> \dot n_{\rm esc}^{\rm bh}$ at $z > 7$. However at $z < 7$ the situation is quite different: the most massive black holes and therefore the most luminous AGN are hosted in massive galaxies. Additionally, the presence of a critical halo mass below which black hole growth is suppressed (see Sec. \ref{delphi}) translates into a critical stellar mass \citep[Fig. 6; ][]{dayal2019}, below which only low-luminosity AGN exist and $\fescbh$ is very low. The fact that both the intrinsic photon production from AGN and $f_{\rm esc}^{\rm bh}$ are very low in low-mass galaxies suppresses the AGN contribution from such galaxies to the escaping photon budget. However, the fact that $\dot n_{\rm int}^{\rm sf} \simeq \dot n_{\rm int}^{\rm bh}$ for high-mass galaxies coupled with an increasing $f_{\rm esc}^{\rm bh}$ value results in black holes {\it dominating} the escaping ionizing radiation rate for galaxies with mass above  a ``transition stellar mass" of $\mstar \gsim 10^{9.6}\, (10^{9.2})\msun$ at $z \sim 6\, (4)$. 

The suppression of black hole growth in low-mass galaxies, advocated from either trying to reconcile seemingly contradictory observational results \citep{2011MNRAS.417.2085V} or from the results of cosmological hydrodynamical simulations \citep{2015MNRAS.452.1502D,bower2017}, modifies the picture compared to early papers that assumed unimpeded growth of massive black holes in small galaxies/halos \citep{volonteri2009}. As noted above, the suppression of black hole contribution from small galaxies/halos, which dominate the mass function at the highest redshifts, is further strengthened by the assumption that $f_{\rm esc}^{\rm bh}$ increases with AGN luminosity. 

The contribution of AGN to reionization was studied using a semi-analytical model also by \cite{2017MNRAS.472.2009Q}. Qualitatively, our results agree with theirs, in the sense that only relatively high-mass black holes are important thus limiting the contribution of AGN to low redshift, and that the AGN contribution to reionization is sub-dominant, of order 10-15\% at $z<6$. The specific assumptions of the models differ, though:  \cite{2017MNRAS.472.2009Q} assume a luminosity-independent obscured fraction, and they do not include a spectral energy distribution that depends on intrinsic black hole properties (mass, accretion rate). In general, models that reproduce the generally accepted UV luminosity functions of galaxies and AGN will all converge to a similar fractional contribution of AGN to reionization. The main reason for the agreement between our results and those of \cite{2017MNRAS.472.2009Q} is that in both models black hole growth is retarded with respect to galaxies, although in different ways. In our model suppression of black hole growth leads to a black hole mass function with a step-like appearance, in their case it is the overall normalization of the mass function that decreases with increasing redshift. In principle, this can be tested observationally through measurements of the relation between black hole and stellar masses in high redshift galaxies. 

As expected from the above discussion, star formation in galaxies dominate $\dot n_{\rm esc}$ for all stellar masses at $z > 7$ although the AGN contribution increases with $\mstar$ as shown in panel (d) of Fig. \ref{fig_ife}. At $z <7 $, however, AGN can start dominating $\dot n_{\rm esc}$ by as much as {\it one order of magnitude} for $\mstar \sim 10^{11}\msun$ galaxies at $z \sim 6$ where black holes can accrete at the Eddington rate. This peak mass shifts to lower $\mstar$ values with decreasing redshift - at $z\sim 4$ AGN in galaxies with masses as low as $\mstar \sim 10^{9.6} \msun$, which can accrete at the Eddington limit, dominate $\dot n_{\rm esc}$ by a factor of 10.

The redshift evolution of the ``transition mass", at which AGN start dominating $\dot n_{\rm esc}$, is shown in panel (e) of the same figure which shows two key trends: firstly, as expected, the transition mass only exists at $z <7$ with stellar radiation dominating $\dot n_{\rm esc}$ at higher-$z$. Secondly, as black holes in galaxies of increasingly lower stellar mass can accrete at the Eddington limit with decreasing redshift (Piana et al., in prep.), the transition mass too decreases with $z$ from $\sim 10^{10.7} \msun$ at $z \sim 6.8$ to $\sim 10^{9.3} \msun$ by $z \sim 4$. In the same panel, we also show a comparison of this transition mass to the observationally-inferred knee of the stellar mass function (${\rm M_*^{knee}}$) which ranges between $10^{10.5}-10^{11}\msun$ at $z \sim 4-7$. While the transition mass is comparable to the knee stellar mass at $z \sim 6.8$, it shows a very rapid decline with decreasing redshift. Indeed, by $z \sim 4$, AGN start dominating $\dot n_{\rm esc}$ from galaxies that are (at least) an order of magnitude less massive compared to the knee mass and in fact the ratio between the escaping \HI ionizing photon rate for AGN and star formation peaks at intermediate galaxy masses. Finally, we note that such a transition mass {\it only exists} in the case that the stellar mass dependence of $\fescsf$ is shallower than $\fescbh$ (see Section \ref{alt_models}). 

We summarise the impact of the the above-noted trends on the production/escape rates of \HI ionizing photons per baryon over a Hubble time in Fig. \ref{fig_bar}. Here the contribution in each galaxy mass range is weighted by its cosmic abundance, via the mass of the host halo - therefore this figure represents the effective contribution of that mass range to the global photon budget. We note that, at any $z$, while $\dot n_{\rm esc}^{\rm sf}$ is just a scaled version of $\dot n_{\rm int}^{\rm sf}$, $\dot n_{\rm esc}^{\rm bh}$ instead evolves based on the luminosity/mass evolution. The key trends emerging are: firstly, at any $z$, whilst the contribution of stars (weighted by the number density) is the highest at intermediate stellar mass galaxies ($10^{7-9}\msun$) at $z \sim 6$, the contribution is essentially mass independent between a stellar mass of $10^{5-8}\msun$ at $z \sim 9$. Although massive galaxies, $\mstar \sim 10^{9}-10^{10} \msun$, have higher production rates of ionizing radiation from both stars and black holes in addition to higher $f_{\rm esc}^{\rm bh}$ values, they are rarer than their low-mass counterparts, which therefore dominate the total emissivity as also shown in the left panel of Fig.~\ref{fig_emm}.  Secondly, AGN only have a contribution at the high stellar mass end ($\mstar \sim 10^{9-10}\msun$) at $z\lsim 9$. Thirdly, as expected from the above discussions, given both the higher values of the intrinsic \HI ionizing photon production rate and $f_{esc}$, AGN dominate the emissivity at the high-mass end ($\mstar \gsim 10^{9}\msun$) at $z \sim 6$. 

Since AGNs are efficient producers of HeII ionizing photons, useful constraints can be obtained on their contribution from the corresponding observations, e.g., \HeII Ly$\alpha$ optical depth at $z \sim 3$ \citep{worseck2016} and the heating of the IGM at $z \lesssim 5$ \citep{becker2011}. A detailed modelling of the \HeII reionization history is beyond the scope of this work. However, we have computed the \HeIII volume filling fraction, $Q_{\mathrm{HeIII}}$, and found that $Q_{\mathrm{HeIII}} \sim 0.4 \, (0.2)$ at $z = 4 \, (5)$, assuming that the escape fraction of \HeII ionizing photons is the same as that of the \HI ionizing photons. While this implies a \HeII reionization earlier than the model of \citet{haardt-madau2012}, it is still within the $2-\sigma$ bounds as allowed by the observations \citep[see, e.g.,][]{mitra2018}.

%{\mv To do: add a note on HeII and temperature.}

% *******************************************
\section{Alternative models}
\label{alt_models}
% *******************************************
Our key result is that the AGN contribution of ionizing photons is subdominant at all galaxy masses at $z>7$. At $z\sim 6-7$ their contribution increases with stellar mass, and at lower redshift it is AGN in intermediate-mass galaxies that produce most ionizing photons (Fig.~\ref{fig_ife}). This results in a ``transition" stellar mass at which AGN overtake the stellar contribution to the escaping ionizing radiation; for stars in galaxies to dominate all the way through in the mass function, either the escape fraction of stellar radiation from galaxies should increase with galaxy mass or that from AGN should decrease, especially at high masses. In our fiducial model, this transition stellar mass decreases with decreasing redshift. Further, star formation in galaxies with mass $<10^9 \msun$ is the main driver of hydrogen reionization. One could argue that this is a consequence of the steep increase of $\fescsf$ at high redshifts, which artificially boosts the contribution of stars in low-mass galaxies and correspondingly reduces the contribution of AGN. In this section we examine the robustness of our results by exploring six different combinations of $f_{\rm esc}^{\rm bh}$ and $\fescsf$ in Sec. \ref{alt_agn} and two different stellar population synthesis models in Sec. \ref{alt_sps} in order to explore the physically plausible parameter space.

% *******************************************
\subsection{Alternative models for AGN and star formation escape fractions}
\label{alt_agn}
% *******************************************
\begin{figure*}
\begin{center}
\center{\includegraphics[scale=1.1]{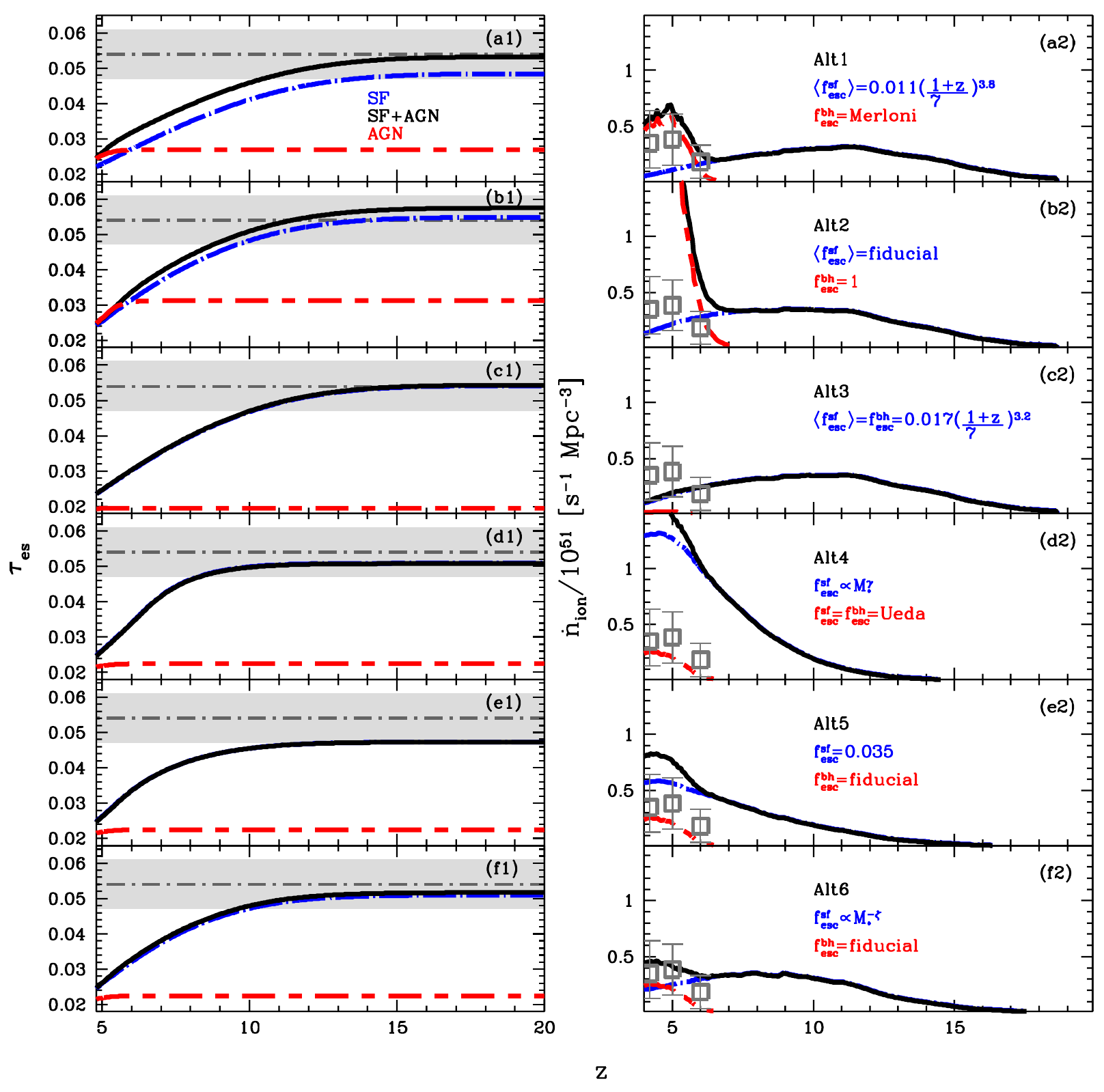}} %tau_nion_extra_cases2.pdf 
\caption{The redshift evolution of the electron scattering optical depth ({\it left column}) and the associated escaping ionizing emissivity ({\it right column}). In the left column, the dot-dashed horizontal line shows the central value for $\tau_{\rm es}$ inferred by the latest Planck results \citep{planck2018} with the gray striped region showing the $1-\sigma$ errors. In the right column, open squares show the observational results (and associated error bars) calculated following the approach of \citet{kfg2012}. In each panel, we show results for star formation+AGN (solid line), star formation (dot-dashed line) and AGN (short-long-dashed line) for the different alternative escape fraction models ({\it Alt1-Alt6}) discussed in Sec. \ref{alt_agn} and summarised in Table \ref{table2a}. The model name and the $f_{esc}$ values used for star formation and AGN are noted in each panel of the right column.  }
\label{fig_extra}
\end{center}
\end{figure*}

\begin{figure*}
\begin{center}
\center{\includegraphics[scale=1.05]{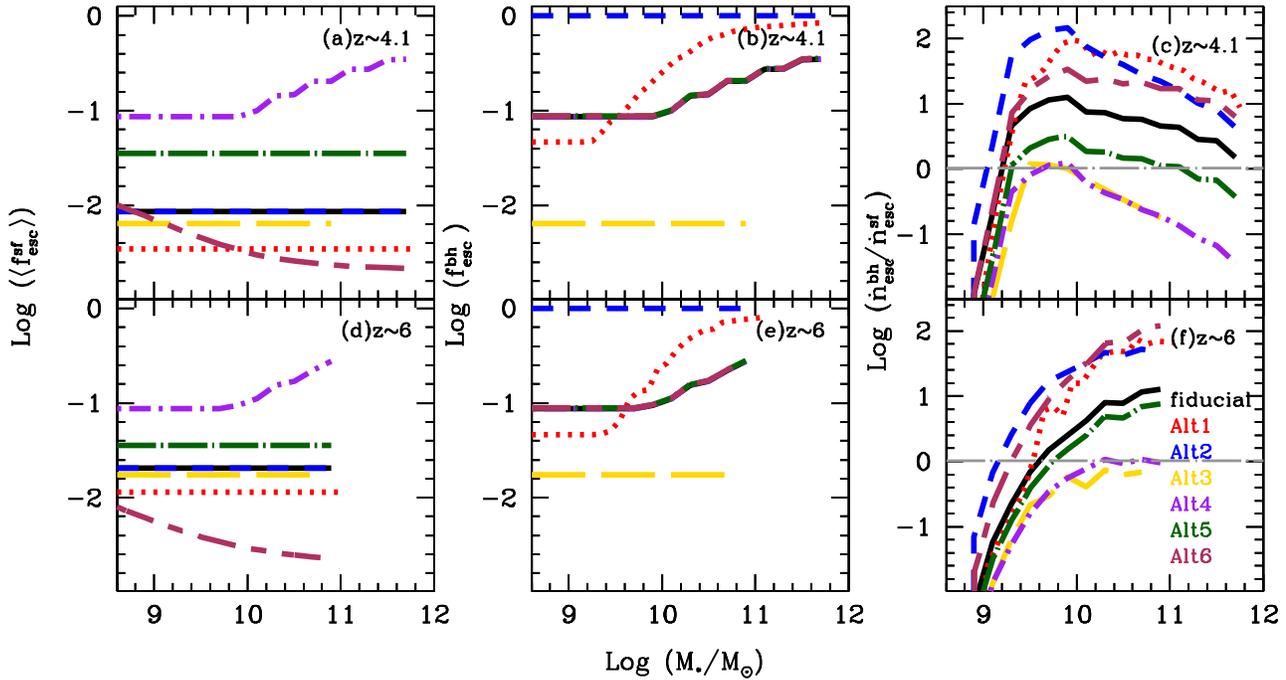}}
\caption{As a function of stellar mass, we show $\fescsf$ (left column), $f_{\rm esc}^{\rm bh}$ ({\it middle column}) and the ratio between the escaping \HI ionizing photon rate for AGN and stars ({\it right column}) for $z \sim 4.1$ (top row) and $z \sim 6$ (bottom row). We show results for the six different alternative escape fraction models ({\it Alt1- Alt6}) discussed in Sec. \ref{alt_agn} and summarised in Table \ref{table2a} and also plot the fiducial model for comparison. In the right-most column, the horizontal line shows a ratio of unity.}
\label{fig_ife_extra}
\end{center}
\end{figure*}

Given that the trends of $\fescsf$ and $\fescbh$ with galaxy properties are still uncertain, both theoretically and observationally,  Fig.~\ref{fig_extra} shows the optical depth and emissivity predicted by the alternative models summarised in Table \ref{table2a}:

\begin{table*}
\begin{center}
\caption{For the alternative models studied in Sec. \ref{alt_agn}, we summarise the model name (column 1), the parameter values for $\fescsf$ (column 2) and $\fescbh$ (column 3), the impact on the ratio $\dot n_{\rm esc}^{\rm bh}/\dot n_{\rm esc}^{\rm sf}$ compared to the fiducial model (column 4) and the impact on the transition mass at which AGN start dominating the escaping \HI ionizing photon production rate compared to the fiducial model (column 5). We note that of models {\it Alt1 - Alt6}, only {\it Alt1}, {\it Alt3} and {\it Alt6} simultaneously fit $\tau_{\rm es}$ \citep{planck2018} and the redshift evolution of the \HI ionizing photon emissivity. We use the {\it fiducial} values of the free parameters for galaxy formation as in Table \ref{table1}.} 
\begin{tabular}{|c|c|c|c|c|}
\hline
Model & $\fescsf$ & $\fescbh$ & $\dot n_{\rm esc}^{\rm bh}/\dot n_{\rm esc}^{\rm sf}$ & Transition ${\rm M_*}$  \\
\hline
{\it Alt1} & $0.017 [(1+z)/7]^{3.8}$  &  \citet{merloni2014} & Increases at all $\rm M_*$ & Almost unchanged\\
{\it Alt2} &  fiducial &  1 & Increases at all $\rm M_*$ & Decreases by 0.2 (0.4 dex) at $z \sim 6\,(4)$\\
{\it Alt3} &  $0.017[(1+z)/7]^{3.2}$ & $0.017 [(1+z)/7]^{3.2}$ & Decreases at all $\rm M_*$ & -\\
{\it Alt4} & ${\rm fiducial}\, \fescbh \propto M_*^\gamma$   & fiducial & Decreases at all $\rm M_*$ & - \\
{\it Alt5} &  0.035 &  \citet{2014ApJ...786..104U} & Decreases  at all $\rm M_*$ for $z \lsim 7.5$ & Increases by 0.1 dex at $z \sim 6-4$\\
{\it Alt6} & $0.1 (f_*^{\rm eff}/f_*^{\rm ej}) \propto M_*^{-\zeta}$  & fiducial & Increases for $\rm M_*\gsim 10^{9.2}\Msun$ & Decreases by 0.3 dex (unchanged) at $z \sim 6$ (4)\\
\hline
\end{tabular}
\label{table2a}
\end{center}
\end{table*}

{\it (i)} In the first model ({\it Alt1}, panels a1 and a2), $f_{\rm esc}^{\rm bh}$ is obtained from the results of \citet{merloni2014}. We fit to the optical depth and emissivity observations to derive $\fescsf=0.017 [(1+z)/7]^{3.8}$. This steep redshift-dependence for the escaping stellar radiation from galaxies (left-most column of Fig. \ref{fig_ife_extra}) is required to off-set the increasing AGN contribution at $z \lsim 5$ which is driven by the higher $f_{\rm esc}^{\rm bh}$ values (compared to the fiducial model) as shown in the middle column of Fig. \ref{fig_ife_extra}. This enhances the ratio $\dot n_{\rm esc}^{\rm bh}/\dot n_{\rm esc}^{\rm sf}$ by more than one order of magnitude compared to the fiducial model at $z <7$ (right-most column of Fig. \ref{fig_ife_extra}). As seen from the same panel, we find that the transition mass remains almost unchanged compared to the fiducial case.

{\it (ii)} In the second model ({\it Alt2}, panels b1 and b2) we keep $\fescsf$ equal to the fiducial value and maximise the escape fraction from AGN by assuming $f_{\rm esc}^{\rm bh}=1$. Driven by such maximal AGN contribution, this model severely over-predicts the emissivity at $z \lsim 5$; the optical depth, being dominated by star formation in galaxies for most of the reionization history, can still be fit within the $1-\sigma$ error bars. As seen from the right-most panel of Fig. \ref{fig_ife_extra}, $\dot n_{\rm esc}^{\rm bh}/\dot n_{\rm esc}^{\rm sf}$ is higher by more than one order of magnitude compared to the fiducial model. Again, a transition stellar mass exists at $z <7 $ and is only slightly lower (by about 0.2-0.4 dex) compared to the fiducial model.

{\it (iii)} In the third model ({\it Alt3}, panels c1 and c2) we consider the same redshift-dependent escape fraction for the ionizing radiation from both stellar radiation and AGN. Here, simultaneously fitting to the optical depth and emissivity values yields an escape fraction that evolves as $\fescsf=f_{\rm esc}^{\rm bh}=0.017 [(1+z)/7]^{3.2}$. The evolution of $\fescsf$ and $f_{\rm esc}^{\rm bh}$ can be seen from the left and middle columns of Fig. \ref{fig_ife_extra}. This model naturally results in a lower AGN contribution to the escaping ionizing radiation at all masses and redshifts as compared to the fiducial model (right most panel of the same figure). Similar to the results of model {\it Alt4} that follows, in this model the AGN ionizing radiation contribution is minimised and only slightly exceeds that from galaxies at $\mstar \sim 10^{9.5-9.8}\msun$ by $z \sim 4$, i.e. stellar radiation dominates the ionizing budget at effectively all masses and redshifts although the AGN contribution still increases with increasing stellar mass. 

\begin{figure*}
\begin{center}
\center{\includegraphics[scale=1.08]{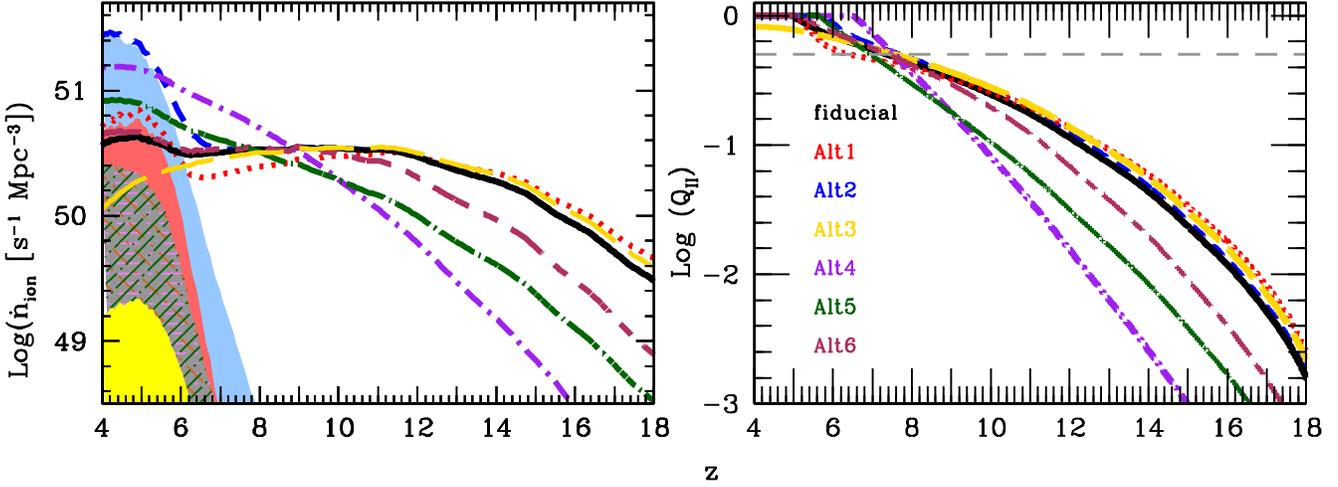}}
\caption{ {\it Left panel}: As a function of redshift, we show the escaping \HI ionizing photon emissivity. The different lines show the emissivity from star formation+AGN while the shaded regions (of the same lighter colour) show the contribution from AGN only. {\it Right panel}: The reionization history, expressed through the redshift evolution of the volume filling fraction of \HII. The horizontal dashed line shows Log$(Q_{II}) = -0.301$, i.e. when reionization is 50\% complete. The different colours in both panels show results for the fiducial and alternative escape fraction models (discussed in Sec. \ref{alt_agn}) as marked in the right panel.}
\label{fig_nioncum}
\end{center}
\end{figure*}

{\it(iv)} In the fourth model ({\it Alt4}, panels d1 and d2) we assume $\fescsf=\fescbh$ using the fiducial $\fescbh$ value from \citet{2014ApJ...786..104U} for galaxies that have a black hole; we use $\fescsf=0$ for galaxies that do not host a black hole. This results in both $\fescsf$ and $\fescbh$ scaling {\it positively} with the stellar mass as shown in the left-most and middle panels of Fig. \ref{fig_ife_extra}. As in the previous model, this identical escape fraction for both stellar radiation and AGN results in stellar radiation dominating the ionizing budget at almost all masses and redshifts; the AGN ionizing radiation contribution only slightly exceeds that from galaxies at $\mstar \sim 10^{10}\msun$ by $z \sim 4$. However, we note that this model over-predicts the emissivity from stellar sources at all redshifts and is unable to simultaneously reproduce both the values of $\tau_{es}$ the the emissivity.

{\it(v)} In the fifth model ({\it Alt5}, panels e1 and e2) we assume a constant $\fescsf = 3.5\%$ and use the fiducial value for $\fescbh$. As seen from the bottom panels of Fig. \ref{fig_extra}, this model is unable to simultaneously reproduce both the values of $\tau_{\rm es}$ and the emissivity. In this model, the value of  
$\fescsf$ is decreased (increased) at $z \gsim 7.5\, ($\lsim 7.5$)$ compared to the {\it fiducial} case as shown in the left panel of Fig. \ref{fig_ife_extra}. Compared to the fiducial model, this results in a lower value of $\dot n_{esc}^{bh}/\dot n_{esc}^{sf}$ by about 0.3 (0.8 dex) at $z \sim 6$ ($z \sim 4.1$) and the transition mass increases negligibly (by $\sim 0.1$ dex) at $z=4-6$. 

{\it(vi)} In the sixth model ({\it Alt6}, panels f1 and f2), while we use the fiducial value for $\fescbh$, we assume that $\fescsf$ scales with the ejected gas fraction such that $\fescsf = f_0 (f_*^{\rm eff}/f_*^{\rm ej})$. This naturally results in $\fescsf$ decreasing with an increasing halo (and stellar) mass. A value of $f_0=0.1$ is required to simultaneously fit both the optical depth and emissivity constraints as shown in the same figure. In this model, the increasing suppression of the star formation rate in low-mass halos due to both supernova and reionization feedback naturally leads to a downturn in the stellar emissivity with decreasing redshift. As shown in Fig. \ref{fig_ife_extra}, in this model the $\fescsf$ values lie below the fiducial one for all $M_* \gsim 10^{8.4}\msun$ at $z \sim 6$. However, by $z \sim 4$, the $\fescsf$ values for the lowest mass halos ($\sim 10^{8.6}\msun$) approach the values for the fiducial model. Compared to the fiducial model, this results in an increasing $\dot n_{esc}^{bh}/\dot n_{esc}^{sf}$ with increasing stellar mass, specially for $M_* \gsim 10^{9.2}\msun$. This naturally leads a transition mass that is lower than that in the fiducial model by about 0.3 dex at $z \sim 6$, whilst being almost identical at $z \sim 4$.

To summarise, the possible range of $\fescsf$ and $\fescbh$ combinations (ranging from redshift-dependent to constant to scaling both positively and negatively with stellar mass) have confirmed our key results: the AGN contribution of ionizing photons is subdominant at all galaxy masses at $z>7$ and increases with stellar mass at $z < 7$. Additionally, we have confirmed the existence of a ``transition" stellar mass (at which AGN overtake the stellar contribution to the escaping ionizing radiation) which decreases with decreasing redshift. Stars dominate all the way through the mass function only when the stellar mass dependence of $\fescsf$ is steeper than $\fescbh$ or if we assume the same $f_{esc}$ values for both star formation and AGN (i.e. the {\it Alt3} and {\it Alt4} models); in this case, naturally, the transition mass no longer exists. 

% ******************************************
\subsection{Alternative stellar population synthesis models}
\label{alt_sps}
% *******************************************
In addition to the fiducial {\it SB99} model, we have considered two other population synthesis models: BPASS binaries \citep[BPB; ][]{eldridge2017} and Starburst99 including stripped binaries \citep[SB99+sb;][]{gotberg2019}. The time evolution of the intrinsic ionizing and UV photons from star formation in the BPB model can be expressed as:

\begin{equation}
\dot n_{\rm int}^{\rm sf}(t)= 10^{47.25} - 2.28 \, {\rm log} \bigg(\frac{t}{2\ \myr} \bigg) + 0.6 \, [{\rm s^{-1}}],
\end{equation}

\begin{table}
\begin{center}
\caption{The parameter values for the $z$-evolution of the escape fraction, $\langle f_{\rm esc}^{\rm sf} \rangle=f_0 [(1+z)/7]^{\beta}$ for different models constrained to simultaneously fit $\tau_{\rm es}$ \citep{planck2018} that combines polarisation, lensing and temperature data, and the redshift evolution of the \HI ionizing photon emissivity (see text). We use the {\it fiducial} value for $\fescbh$ and the same values of the free parameters for galaxy formation as in Table \ref{table1}.}
\begin{tabular}{|c|c|c|c|}
\hline
SPS Model & Sources & $f_0 \times 100$ & $\beta$ \\
\hline
SB99 & SF &  2.0 & 2.8 \\
SB99 & SF+AGN & 1.85  & 2.8 \\
BPB & SF & 0.46 & 2.8 \\
BPB & SF+AGN & 0.43 & 2.8\\
SB99+sb & SF & 1.7  & 2.8\\
SB99+sb & SF+AGN & 1.6 & 2.8\\
\hline
\end{tabular}
\label{table2}
\end{center}
\end{table}

\begin{equation}
L_{\rm UV} (t)= 10^{33.0} - 1.2\, {\rm log} \bigg(\frac{t}{2\ \myr} \bigg) + 0.5 \, [{\rm erg\, s^{-1} \, \AA^{-1}}].
\label{uv_bpb}
\end{equation}
\noindent
In the SB99+sb model, these quantities evolve as:
\begin{equation}
 \dot n_{\rm int}^{\rm sf}(t) = 10^{46.7} - 2.3\, {\rm log} \bigg(\frac{t}{2\ \myr}\bigg)  \, [{\rm s^{-1}}],
\end{equation}
\noindent
\begin{equation}
L_{\rm UV}(t) = 10^{33.01} - 1.3\, {\rm log} \bigg(\frac{t}{2\ \myr}\bigg) + 0.49 \, [{\rm erg\, s^{-1} \, \AA^{-1}}] 
\label{uv_sbpsb}
\end{equation}
\noindent
The rest-frame UV luminosity has almost the same normalisation and time-evolution in all three models (SB99, BPB, SB99+sb) resulting in the same UV LFs. However, as seen from Eqns. \ref{uv_sb99}, \ref{uv_bpb} and \ref{uv_sbpsb}, the slope of the time evolution of  $\dot n_{\rm int}$ is much shallower in the BPB and SB99+sb models compared to the fiducial (SB99) model. We re-tune $\fescsf$ for each of these models to match to the reionization data ($\tau_{\rm es}$ and the emissivity) using the fiducial $f_{\rm esc}^{\rm bh}$ values, the results of which are summarised in Table \ref{table2}. As seen, while the slope of the redshift dependence of $\fescsf$ remains unchanged ($\beta=2.8$), the normalisation ($f_0$) is the lowest for the BPB model as compared to SB99 by a factor 4.6; the SB99 and SB99+sb models on the other hand only differ by a factor 1.17. Finally, the lower $\fescsf$ values compensate for a higher intrinsic production rate to result in the same $\dot n_{\rm esc}^{\rm sf}$ value as a function of $\mstar$. These different stellar populations, therefore, have no bearing on our result regarding the relative AGN/starlight contribution to the ionizing radiation for different galaxy stellar masses.

\begin{figure*}
\begin{center}
\center{\includegraphics[scale=0.75]{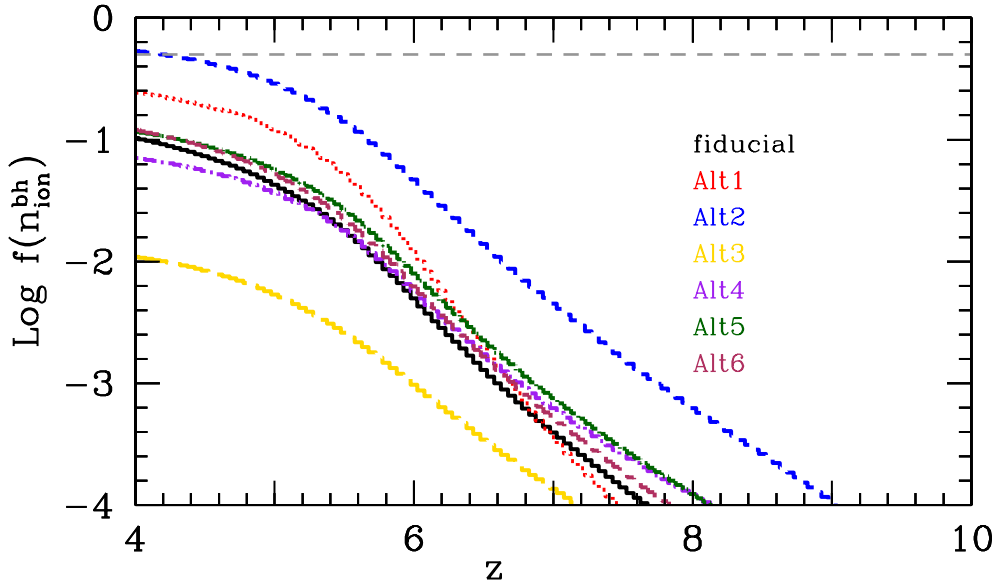}}
\caption{The cumulative fraction of ionizing photons contributed by AGN as a function of redshift; the horizontal short-dashed line shows the 50\% contribution to the cumulative ionizing emissivity for the various models discussed in this work (see Sec. \ref{alt_agn} for details), as marked.  }
\label{fig_qii}
\end{center}
\end{figure*}

% *******************************************
\section{Reionization history and the cumulative AGN contribution}
% *******************************************
We start with a recap of the total (star formation+AGN) ionizing emissivity for all the different models considered in this work in (the left-panel of) Fig. \ref{fig_nioncum}. In all models, the ionizing emissivity from star formation dominates at $z > 6$ and is virtually indistinguishable for all the models (fiducial, {\it Alt1, Alt2} and {\it Alt3}) that use a redshift dependent $\fescsf$ value. The redshift evolution of the emissivity is the steepest for the {\it Alt4} model where $\fescsf \propto M_*$. With its constant value of $\fescsf =0.035$, model {\it Alt5} shows the shallowest slope. Given its lower $\fescsf$ values for all stellar masses at high-redshifts, the {\it Alt6} model naturally shows a lower ionizing emissivity compared to fiducial; the stellar emissivity from the {\it Alt6} model converges to the fiducial one by $z \sim 9$ as a result of the decreasing $\fescsf$ values for the latter. As expected, the AGN contribution is the lowest for the model {\it Alt3} where $\fescsf=\fescbh=$ a decreasing function of redshift (as shown in the same panel). It then increases by a factor of 3 from the fiducial case to the {\it Alt1} case and reaches its maximum for the {\it Alt2} case where $\fescbh=1$.  

We then discuss reionization history, expressed through the redshift evolution of the volume filling fraction of ionized hydrogen ($Q_{II}$), as shown in (the right-panel of) Fig. \ref{fig_nioncum}. Interestingly, despite the range and trends used for $\fescsf$ and $\fescbh$, reionization is 50\% complete in all cases in the very narrow redshift range of $z \sim 6.6-7.6$. Further, we find an end redshift of reionization value of $z_{re} \sim 5-6.5$ in all the models studied here except {\it Alt 3}. In this model, the decrease in the star formation emissivity (driven by the decrease of $\fescsf$) with decreasing redshift is not compensated by an increasing AGN contribution as in the other models; as a result, reionization does not finish even by $z \sim 4$. Given that star formation in low-mass halos is the key driver of reionization, it is not surprising to see that reionization finishes first ($z_{re} \sim 6.5$) in the {\it Alt4} model that has the largest value of $\fescsf$. Models {\it Alt2} and {\it Alt5} show a similar $z_{re} \sim 5.8$ driven by an increasing contribution from star formation and AGN, respectively. Finally, given their lower values of the total ionizing emissivity at $z \lsim 7$, reionization ends at $z_{re} \sim 5$ in the fiducial, {\it Alt1} and {\it Alt6} models.

Finally, we show the AGN contribution to the cumulative ionizing emissivity as a function of redshift in Fig. \ref{fig_qii}. As seen, AGN contribute at most 1\% of the total escaping ionizing photon rate by $z \sim 4$ in the {\it Alt3} model. This increases to $\sim 10\%$ of the total ionizing emissivity for the fiducial and {\it Alt4-Alt6} cases. Compared to the fiducial case, the higher $\fescbh$ in the {\it Alt1} case results in an AGN contribution as high as $25\%$ by $z \sim 4$. Finally, the {\it Alt2} case ($\fescbh=1$) provides the upper limit to the AGN contribution. Here, AGN contribute as much as galaxies to the cumulative emissivity by $z \sim 4.4$. 

In addition to the fiducial model, only {\it Alt1}, {\it Alt3} and {\it Alt6} are able to simultaneously reproduce the emissivity and optical depth constraints. However, as seen above, the {\it Alt3} model does not have enough ionizing photons to finish the process of reionization. This leaves us with three physically plausible models - the fiducial one, {\it Alt1} and {\it Alt6}. In these, the AGN contribution to the total emissivity is sub-dominant at all $z$; AGN contribute about $0.5-1\%$ to the cumulative ionizing emissivity by $z \sim 6$ that increases to $10-25\%$ by $z=4$. 

% *******************************************
\section{Conclusions}
% *******************************************
\label{sec:conclusions}

In this paper, we have studied the contribution of AGN to hydrogen reionization. Our model includes a delayed growth of black holes in galaxies via suppression of black hole accretion in low-mass galaxies, caused by supernova feedback. Furthermore, in our model each accreting black hole has a spectral energy distribution that depends on the black hole mass and accretion rate. Given that the escape fractions for both star formation and AGN remain poorly understood, we have explored a wide range of combinations for these (ranging from redshift-dependent to constant to scaling both positively and negatively with stellar mass). Using these models, we find the following key results:

\begin{itemize}

\item{The {\it intrinsic} production rate of ionizing photons for both star formation and AGN scales positively with stellar mass with star formation dominating at all masses and redshifts.}

\item{Irrespective of the escape fraction values used, the AGN contribution to the {\it escaping} ionizing photons is always sub-dominant at all galaxy masses at $z >7$. In the case that the stellar mass dependence of $\fescsf$ is shallower than $\fescbh$, at $z<7$ a ``transition" stellar mass exists above which AGN dominate the escaping ionizing photon production rate. This transition stellar mass decreases with redshift from being equal to the knee of the stellar mass function at $z \sim 6.8$ to being an order of magnitude less than the knee by $z=4$.}

%If one assumes (as we did in the fiducial model and models {\it Alt1, Alt4} and {\it Alt5} in Sec. \ref{alt_agn}) that $f_{\rm esc}^{\rm bh}$ scales with the fraction of unabsorbed AGN, then $f_{\rm esc}^{\rm bh}$ is an increasing function of stellar mass. Massive galaxies, $\mstar \sim 10^{9}-10^{10} \msun$, have therefore a higher production of ionizing radiation, from both stars and black holes, and they also have higher $f_{\rm esc}^{\rm bh}$.}

\item{Overall, the ionizing budget is dominated by stellar radiation from low-mass ($M_*<10^9 \msun$) galaxies down to $z \gsim 6$ in all models. In the fiducial model, at $z=6$ AGN and stars in $M_*>10^9 \msun$ contribute equally to the ionizing budget ($\sim15\%$ of the total). However at $z<5.5$, the AGN contribution (driven by $M_{bh}>10^6 \msun$ black holes in $M_* \gsim 10^9\msun$ galaxies) overtakes that from star formation in $M_*<10^9 \msun$ galaxies. The contribution from star formation in high-mass ($M_*>10^9\msun$) galaxies is sub-dominant at all redshifts, reaching a maximum value of $20\%$ of the total ionizing budget at $z \lsim 6$. }

\item{Different stellar population synthesis models (SB99, BPB, SB99+sb) have no bearing on our result regarding the relative AGN/starlight contribution to the ionizing radiation for different galaxy stellar masses.}

\item{For all models that match the observed reionization constraints (electron scattering optical depth and the ionizing emissivity) and where reionization finishes by $z \sim 5$, AGN can contribute as much as $~50-83\%$ of the emissivity at $z =5$. However, AGN only contribute $0.5-1\%$ to the {\it cumulative} ionizing emissivity by $z \sim 6$ that increases to $10-25\%$ by $z=4$.}

\end{itemize}

% ***************************************************************************
%\vspace{-0.8cm}
\section*{Acknowledgments} 
% ****************************************************************************
PD acknowledges support from the European Research Council's starting grant ERC StG-717001 (``DELPHI"), from the NWO grant 016.VIDI.189.162 (``ODIN") and the European Commission's and University of Groningen's CO-FUND Rosalind Franklin program. MV and MT acknowledge funding from the European Research Council under the European Community's Seventh Framework Programme (FP7/2007-2013 Grant Agreement no.\ 614199, project ``BLACK''). TRC acknowledges support from the Associateship Scheme of ICTP. MV and RS acknowledge support from the Amaldi Research Center funded by the MIUR program ``Dipartimento di Eccellenza" (CUP:B81I18001170001). MT is supported by Deutsche Forschungsgemeinschaft (DFG, German Research Foundation) under Germany's Excellence Strategy EXC-2181/1 - 390900948 (the Heidelberg STRUCTURES Cluster of Excellence). HA is supported by the Centre National d'Etudes Spatiales (CNES). MH acknowledges financial support from the Carlsberg Foundation via a Semper Ardens grant (CF15-0384). Finally, PD and MV thank La Sapienza for their hospitality, where the bulk of this work was carried out and thank D. Stark and P. Oesch for their insightful comments.

% **************************************************************************

%%%%%%%%%%%%%%%%%%%%%%%%%%%%%%
%\vspace{-0.8cm}
\bibliographystyle{mn2e}
\bibliography{mainbib,mybib}

\appendix
\section{Ionizing properties as a function of black hole properties}

\begin{figure}
\begin{center}
\center{\includegraphics[scale=0.4]{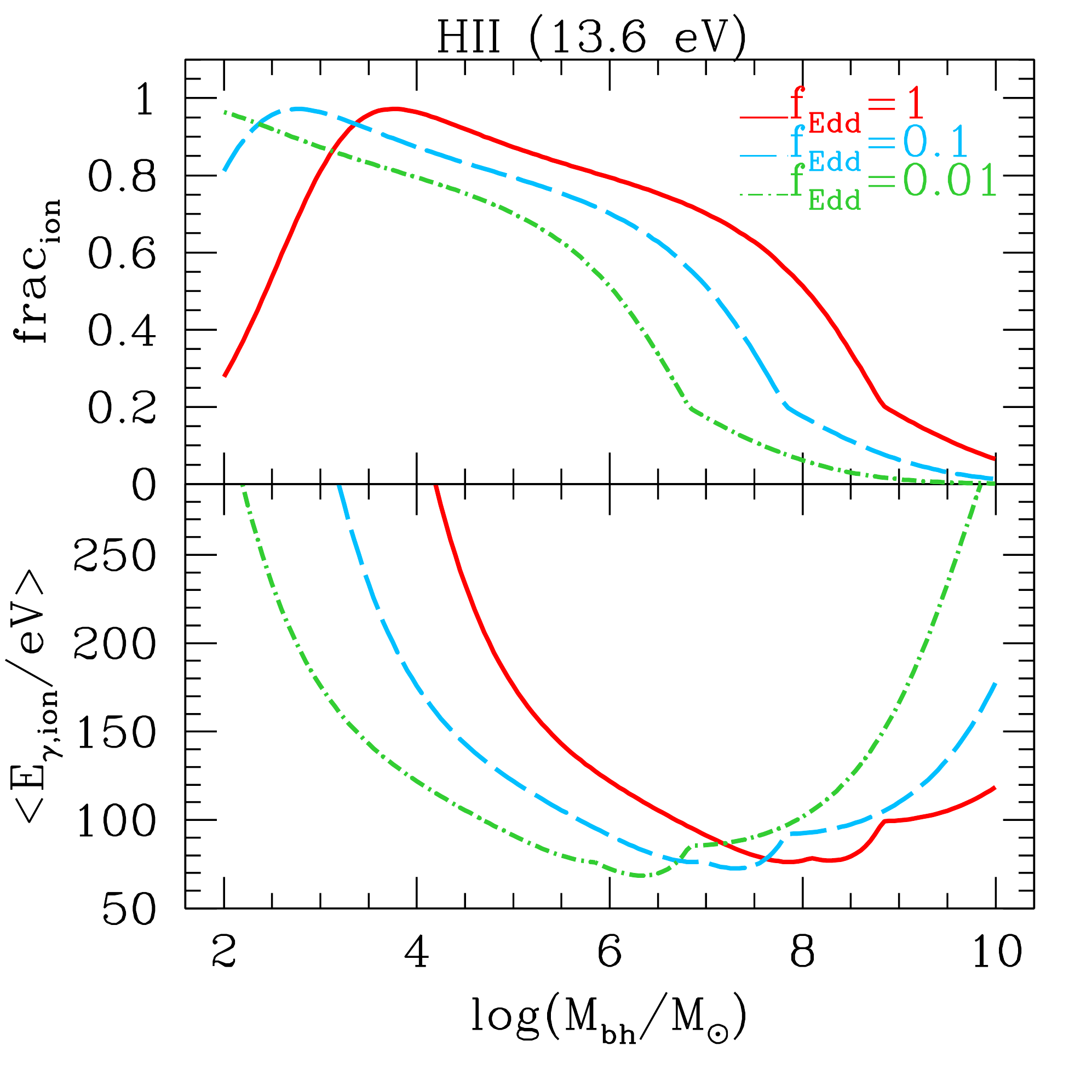}} 
\center{\includegraphics[scale=0.4]{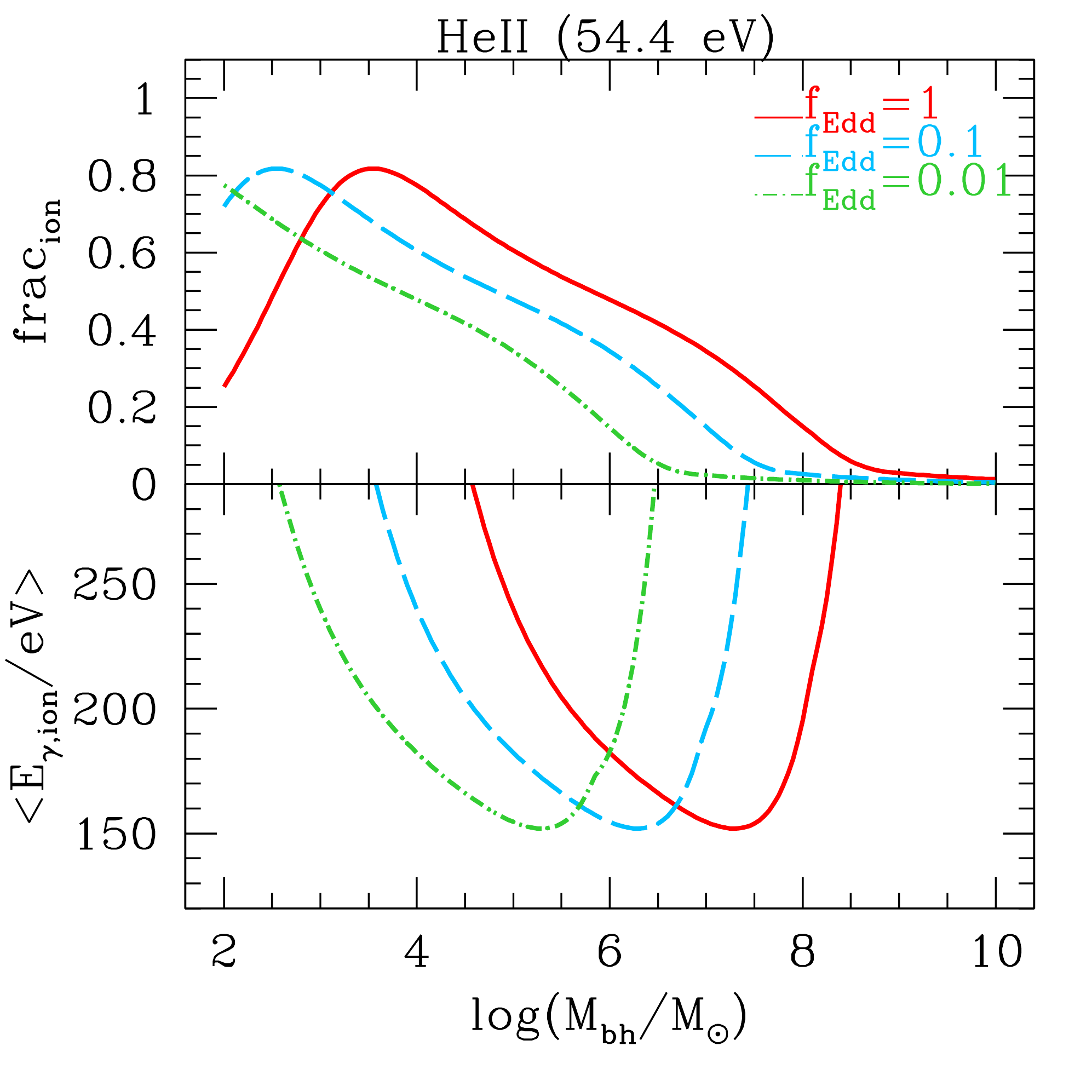}} 
\caption{As a function of black hole mass, the panels (top to bottom) show the fraction of luminosity emitted in photons above 13.6~eV and the mean energy of such photons,  the fraction of luminosity emitted in photons above 54.4~eV and the mean energy of such photons. Solid: for a black hole at the Eddington luminosity; dashed: for a black hole at 10\% of the Eddington luminosity; dot-dashed: for a black hole at 1\% of the Eddington luminosity.}
\label{fig_app}
\end{center}
\end{figure}

\label{lastpage} 
\end{document}